\documentclass[11pt]{article}
 \usepackage{epsfig,amscd,amssymb}
\usepackage{amssymb,amsfonts}
\usepackage{epstopdf}
\usepackage{graphicx}
\usepackage{psfrag}
\usepackage{subfigure}
\usepackage{epsf}
\usepackage{epsfig}
\usepackage{amssymb}
\usepackage{amsmath}
\def\bea{\begin{eqnarray}}
\def\eea{\end{eqnarray}}
\def\be{\begin{equation}}
\def\ee{\end{equation}}

\newcommand{\eps}{\epsilon}

\begin{document}

\def\wgta#1#2#3#4{\hbox{\rlap{\lower.35cm\hbox{$#1$}}
\hskip.2cm\rlap{\raise.25cm\hbox{$#2$}}
\rlap{\vrule width1.3cm height.4pt}
\hskip.55cm\rlap{\lower.6cm\hbox{\vrule width.4pt height1.2cm}}
\hskip.15cm
\rlap{\raise.25cm\hbox{$#3$}}\hskip.25cm\lower.35cm\hbox{$#4$}\hskip.6cm}}

\def\wgtb#1#2#3#4{\hbox{\rlap{\raise.25cm\hbox{$#2$}}
\hskip.2cm\rlap{\lower.35cm\hbox{$#1$}}
\rlap{\vrule width1.3cm height.4pt}
\hskip.55cm\rlap{\lower.6cm\hbox{\vrule width.4pt height1.2cm}}
\hskip.15cm
\rlap{\lower.35cm\hbox{$#4$}}\hskip.25cm\raise.25cm\hbox{$#3$}\hskip.6cm}}

\def\begeqar{\begin{eqnarray}}
\def\endeqar{\end{eqnarray}}

%
%
%
%
\title{ SCATTERING  AND DUALITY IN THE 2 DIMENSIONAL
 $OSP(2|2)$  GROSS NEVEU AND SIGMA MODELS  }

\author{  Hubert Saleur$^{a,b}$ and Bal\'azs Pozsgay$^{c}$\\
\smallskip\\
$^{a}$ Service de Physique Th\'eorique\\
CEA Saclay\\
Gif Sur Yvette, 91191\\
France\\
\smallskip\\
$^{b}$ Department of Physics and Astronomy\\
University of Southern California\\
Los Angeles, CA 90089\\
USA\\
\smallskip\\
$^{c}$ HAS Research Group for Theoretical Physics\\
H-1117 Budapest\\
P\'azm\'any P\'eter s\'et\'any 1/A\\
Hungary\\
\smallskip\\
}

\maketitle

\begin{abstract}
We write the thermodynamic Bethe ansatz  for the massive $OSp(2|2)$ Gross Neveu and sigma models. 
We find evidence that  the  GN S matrix proposed by Bassi and Leclair  \cite{BassiLeclair}  is the correct one. We determine features 
 of the sigma model S matrix,  which seem highly unconventional; we  conjecture in particular a relation  between this 
sigma model and the complex sine-Gordon model at a particular value of the coupling. 
We uncover an intriguing duality between the $OSp(2|2)$ GN (resp. sigma) model on the one hand,  and the $SO(4)$ sigma (resp.   GN model) on the other, somewhat generalizing to the massive case recent  results on $OSp(4|2)$. Finally, we write the TBA  for the (SUSY version of the) flow into the random bond Ising model  proposed by Cabra et al.  \cite{Giuseppe}, and conclude that their S matrix  cannot be correct.

\end{abstract}

\section{Introduction}

The study of $1+1$  quantum field theories with supergroup symmetry is a lively and difficult topic, central to several key areas of modern physics, in particular the solution of critical points in non interacting disordered systems, and the AdS/CFT conjecture. 

It is natural to try to make progress by focussing on  integrable theories. By analogy with the ordinary case, one can expect the simplest possible situations to be encountered \cite{Birgit1} in the $OSp(n|2m)$ Gross Neveu and sigma models. Yet, apart from the special case of  $OSp(1|2)$ which behaves much like an ordinary group \cite{Tsuboi1,Birgit2}, only little progress has been accomplished, largely for what seems to be technical reasons. Another exception is the case of $OSp(4|2)$, where, among other fascinating properties \cite{Mann}, a striking duality between the GN and sigma model was uncovered \cite{CanduSaleur,Volkeretal}. 

The case of $OSp(2|2)$ is particularly tentalizing, since it is the `simplest' after $OSp(1|2)$, and plays a major role in disordered systems. Yet, even here, significant progress has only occurred up to now in the critical case, where WZW theories have been solved \cite{VolkerandI}, and their relation with spin chains understood \cite{EFS,GalleasMartins}. Our main goal in this paper is to understand the {\sl massive theories}: the GN and $OSp(2/2)/OSp(1/2)$ `supersphere' sigma models. 

To proceed, we recall some general features in the ordinary  $O(N)$ bosonic case. The actions are well known \cite{ZZ}. 
The S matrices  involve generically particles in the \ (vector) representation. Introducing the useful graphical representation of invariant tensors
 \begin{figure}[ht]
\centering
\includegraphics[scale=.5]{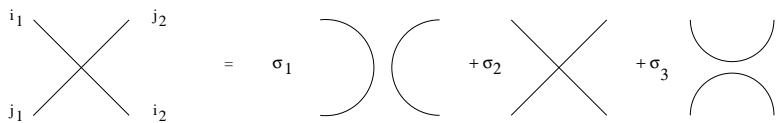}
  \caption{
Graphical representation of the invariant tensors appearing
  in the $S$ matrix.}
  \label{figb1}
 \end{figure}
 there are 
generically two known 
models whose scattering  matrix for the vector representation has the 
form  in figure \ref{figb1}, with none of the $\sigma_i$'s vanishing. They 
are given by 
\begin{eqnarray}
\sigma_1&=&-{2i\pi\over (N-2)(i\pi-\theta)}~\sigma_2\nonumber\\
\sigma_3&=&-{2i\pi\over (N-2)\theta}~\sigma_2\label{mainii}
\end{eqnarray}
with two possible choices for $\sigma_2$:
\begin{equation}
\sigma_2^{\pm}(\theta)={\Gamma\left(1-{\theta\over 2i\pi}\right)\over 
\Gamma\left({\theta\over 2i\pi}\right)}
{\Gamma\left({1\over 2}+{\theta\over 2i\pi}\right)\over 
\Gamma\left({1\over 2}-{\theta\over 2i\pi}\right)}
{\Gamma\left(\pm{1\over N-2}+{\theta\over 2i\pi}\right)\over  
\Gamma\left(1\pm{1\over N-2}-{\theta\over 2i\pi}\right)}
{\Gamma\left({1\over 2}\pm{1\over N-2}-{\theta\over 
2i\pi}\right)\over \Gamma\left({1\over 2}\pm{1\over N-2}+{\theta\over 
2i\pi}\right)}\label{mainiii}
\end{equation}
The factor $\sigma_2^+$ does not have  poles in the physical strip 
for $N\geq 0$, and the corresponding S matrix for $N\geq 3$  is 
believed to describe the $O(N)/O(N-1)$ 
sphere ($S^{N-1}$) sigma model. The factor $\sigma_2^-$ does not have 
poles in the physical strip for $N\leq 4$. For $N>4$, it describes 
the 
scattering of vector particles in $O(N)$ Gross Neveu model. Recall 
that for $N=3,4$ the vector particles in the GN model are unstable and disappear from 
the spectrum, that contains only kinks. Some of these features are 
illustrated for convenience in figure 2.

 \begin{figure}[ht]
\centering
\includegraphics[scale=.5]{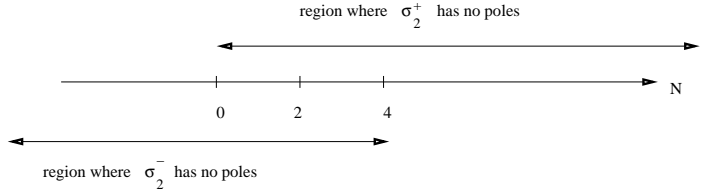}
  \caption{Pole structure of $\sigma_2$ as a function of $N$. Except for a finite region at small $N$,  the generic S matrix with $\sigma_2^+$  describes the sigma model and the one with $\sigma_2^-$ the GN model. }
\label{figb2}
 \end{figure}

Note that at vanishing rapidity, the scattering matrix reduces to 
$\check{S}(\theta=0)=\mp I$. This is in agreement with the 
fundamental particles being bosons in the sigma model , and fermions 
in the Gross Neveu model \cite{KT}. 

It was argued in \cite{Birgit1} that for $N$ negative and $|N|$ large enough, a similar S matrix - where now the invariant tensors have to be interpreted within $OSp(m|2n)$ representation theory - still describes the GN and sigma models provided one sets $N=m-2n$. The detailed calculations performed in \cite{Birgit1} show that one  the roles of the two models {\sl remains the same}, that is, $\sigma_+$ still corresponds to the sigma model and $\sigma_-$ to the GN model. 

This has several interesting consequences. One is, that the GN model exhibits bound states while the sigma model does not, for $N$ large, ie $m>>2n$; meanwhile, the sigma model does then exhibit bound states - and the GN none - for $-N$ large, ie $m<<2n$. Where exactly these behaviours may be modified at small $N$ is a partly open question. We emphasize that, although the S matrices are presumably known in the generic case, no systematic check has been performed, as writing the TBA has proven, up to now, too cumbersome.

A crucial observation based again on \cite{Birgit1} and also \cite{Birgit2} is that the massive continuum limit of the integrable lattice model based on the fundamental of $OSp(m|2n)$  always exhibits bound states. Therefore it necessarily corresponds to the GN model for $N>2$ and the sigma model for $N<2$ (including $N$ negative of course). The case $N=2$ being at the border suggests that the sigma and the GN model are then   dual of each of other. This  is of course well known for $O(2)$, and has recently been established, for a certain choice of  boundary conditions, in the orthosymplectic case \cite{CanduSaleur}, \cite{Volkeretal}. Note that the S matrices for the sigma and GN model at $N=2$ also coincide. They  describe a special point on the critical lines of either (dual) model.

The case $N=0$ ($OSp(2n|2n)$) is the next case that seems really worth studying. Indeed,  right at $N=0$, 
$\sigma_2^+=\sigma_2^-$ (and the foregoing S matrix coincides with the one in  \cite{BassiLeclair}), while it is clear that the GN and sigma models are different (see below for detailed actions etc). Something is thus missing from the general picture. On the other hand, the TBA should not be too complicated. It is particularly tractable for a special choice of Dynkin diagram first studied in \cite{PF}.

The $OSp(2|2)$ has also the potential to exhibit interesting duality properties.  To see this, consider another  - a priori more anecdotical - remark about  the beta functions of the sigma and GN  models
\footnote{Note that the $\beta$ functions  do not exhibit any singularity when going from positive to negative values of $N$. The beta function of say the GN model for $N$ negative is simply obtained from the one for the usual GN model by substituting a negative value of $N$
\cite{Balogetal}. Similarly, the $\beta$ functions are not affected by the ``twisting'' of the fermion boundary conditions.}. In an expansion in the coupling constant  the first two non trivial coefficients  obey:
\begin{eqnarray}
 {\beta_2\over \beta_1^2}={1\over N-2}\hbox{ sigma model}\nonumber\\
 {\beta_2\over \beta_1^2}=-{1\over N-2}\hbox{ GN model}
 \end{eqnarray}
These two results are exchanged under $N-2\rightarrow 2-N$. This suggests there might be some relationship  between the $O(N)$ GN model and $O(4-N)$ sigma models, where an $OSp$ supergroup has to be used whenever necessary.
 
 Of course, these two models cannot be related in general:  among other things, their  UV central charges don't match for arbitrary N. 
 Nevertheless, as observed in \cite{Birgit1}, some of the S matrix elements match exactly under this transformation. One has indeed,
 \begin{equation}
\left[ \sigma_2^+-\sigma_3^+ \right](N)=\left[\sigma_2^-+\sigma_3^-\right](4-N)
 \end{equation}
 This implies that, at zero temperature and with the proper  applied field $h$ , the ground state energies of the $O(N)$ GN model and the one of the $O(4-N)$ sigma model do indeed coincide as functions of $h/M$ (and in particular, their zero field energies - the regularized version of the ground state energy as a function of the mass $M$ - coincide). 
 
Now, if we take the $OSp(2|2)$ GN  ($N=0$) model we have $c=0,c_{\hbox{eff}}=3$ \footnote{We denote in general by $c_{\hbox{eff}}$ the central charge in the (Neveu Schwarz)  sector where the fermions have antiperiodic boundary conditions (on the cylinder and on the plane, since their dimension is integer.}, while for the $SO(4)$ sigma model we have $c=3$ (no fermions in this case). So some identificatiom might be possible. Similarly, if we take the $OSP(2|2)$ sigma model we have $c=-1,c_{\hbox{eff}}=2$ while for the $SO(4)$ GN model, which is a sum of two independent sine Gordon models at the marginal point,  $c=2$, again allowing a potential identification. This pattern generalizes immediately to the case of $OSp(2n|2n)$ and $OSp(2n|2n-4)$:
\begin{eqnarray}
OSp(2n|2n) \hbox{ GN model }: c=0,c_{\hbox{eff}}=3n\nonumber\\
OSp(2n+2|2n-2) \hbox{Sigma model }: c=3,~c_{\hbox{eff}}=3n\nonumber\\
OSp(2n|2n) \hbox{Sigma model}: c=-1,~c_{\hbox{eff}}=3n-1\nonumber\\
OSp(2n+2|2n-2)\hbox{GN model }: c=2,~c_{\hbox{eff}}=3n-1
\end{eqnarray}
One of the questions we will investigate in this paper is the potential duality between these two families of models for the simplest case $n=1$.

Our strategy will be based on non perturbative calculations of the free energies at finite temperature of all models using an S matrix and thermodynamic Bethe ansatz approach. 

Because of the common underlying $osp(2|2)\equiv sl(2|1)$ symmetry, the different Bethe ans\"atze involved have a common structure of roots and basic equations, which coincide with those for the integrable spin chains. We thus discuss these common features first.

    %
%
    %

\section{Solving the monodromy problem: the Bethe ansatz  for integrable  SL(2$|$1)  vertex models and spin chains}

\subsection{The Bethe equations in the fermionic grading}

We start with the Bethe equations for the $sl(2/1)$ spin chain built out of alternating $3$ 
and $\bar{3}$  representations, that is the Hilbert space is $3^{\otimes L}\otimes \bar{3}^{\otimes L}$.  There are different Bethe ans\"atze available, and as often,  chosing the right one can be a big help. We take the Bethe ansatz based on a choice of purely fermionic simple roots \cite{Doikou} for the $sl(2|1)$ algebra.

 \begin{figure}[ht]
 \begin{center}
 \noindent
 \includegraphics[]{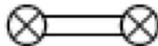}
 \end{center}
 \caption{Dynkin diagram leading to our Bethe equations
  }
 \end{figure}

Recall that the Bethe equations 
read  then \cite{PF,EFS}
\begin{eqnarray}
    \left({u_{j}+i\over 
    u_{j}-i}\right)^{L}=\prod_{\beta=1}^{M}{u_{j}-\gamma_{\beta}+i\over
    u_{j}-\gamma_{\beta}-i},~~~j=1,\ldots,N\nonumber\\
    \left({\gamma_{\alpha}+i\over 
       \gamma_{\alpha}-i}\right)^{L}=\prod_{k=1}^{N}{\gamma_{\alpha}-u_{k}+i\over
       \gamma_{\alpha}-u_{k}-i},~~~\alpha=1,\ldots,M\label{baei}
   \end{eqnarray}
with the quantum numbers $B={N-M\over 
2}$and  $J^{3}=L-{M+N\over 2}$. The corresponding energy, for the regime of interest here, is 
is
\begin{equation}
    E\propto -\left(\sum_{k=1}^{N}{1\over u_{k}^{2}+1}
    +\sum_{\beta=1}^{M}{1\over 
    \gamma_{\beta}^{2}+1}\right)\label{energy}
\end{equation}
A   crucial observation \cite{EFS} which allows one to draw some conclusions 
quickly 
is that  this system of equations admits 
a particular solution made of sets such that $N=M$ and 
$u_{j}=\gamma_{\beta}$. The roots then have to satisfy
\begin{equation}
    \left({u_{j}+i\over 
       u_{j}-i}\right)^{L}=\prod_{k=1}^{n}{u_{j}-u_{k}+i\over
       u_{j}-u_{k}-i}
\end{equation}
a system of equations that is identical with the Bethe equations for 
the spin 1 antiferromagnetic XXX chain (we often refer to the latter 
simply as XXX, or the Takhtajan-Babujian \cite{TB}, chain). Notice however
that the dynamics 
is such that the gaps in our system will be {\sl twice} the gaps in 
the latter chain: sound velocities have to be adjusted appropriately. Often, we will see that our results are in some 
sense a `doubling' of those for this XXX chain.

The argument extends  to a chain with 
 alternating fully symmetric and antisymmetric representations, 
 represented by  Young diagrams with $p$ boxes . The 
 bare Bethe ansatz equations then look exactly as (\ref{baei}) but in 
 the left hand side, the factors $i$ are replaced by $pi$, and 
 similarly for the energy (\ref{energy}) the $1$ in the denominator is replaced by $p^2$. This time, the symmetric sector $N=M$,  $u_j=\gamma_\beta$, is related with the  XXX chain of (integer) spin $p$. 
 

 The argument also works for the model in the fundamental 
 representation of $osp(2/2)$. In that case indeed, the bare Bethe 
 ansatz equations and the energy still read as (\ref{baei},\ref{energy}) but with the factor $i$ 
 replaced by $i/2$ (ie, formally, $p=1/2$ in the foregoing 
 discussion): 
\begin{eqnarray}
    \left({u_{j}+i/2\over 
    u_{j}-i/2}\right)^{L}&=&\prod_{\beta=1}^{M}{u_{j}-\gamma_{\beta}+i\over
    u_{j}-\gamma_{\beta}-i},~~~j=1,\ldots,N\nonumber\\
    \left({\gamma_{\alpha}+i/2\over 
       \gamma_{\alpha}-i/2}\right)^{L}&=&\prod_{k=1}^{N}{\gamma_{\alpha}-u_{k}+i\over
       \gamma_{\alpha}-u_{k}-i},~~~\alpha=1,\ldots,M,~~~\hbox{fundamental of $OSp(2|2)$}\label{baeii}
   \end{eqnarray}
This time the symmetric sector is related with the spin $1/2$ chain.  Note that, while the equations (\ref{baei}) hold for a 
 system made of $L$ representations $3$ {\sl and} $L$ 
 representations $\bar{3}$, when it comes to the fundamental of 
 $osp(2/2)$,  (\ref{baeii}) holds for $L$ representations $4$.

\subsection{The solutions for large chains}

Like for the XXX chain, the types of solutions  of the Bethe equations do not depend on the source terms, and can be discussed in full generality \footnote{The following is borrowed from \cite{EFS} with only small modifications.}. The standard way to classify the solutions  is to
 consider configurations of spectral parameters that sit on poles of
 the bare scattering kernels (the right-hand-sides in
 (\ref{baei},\ref{baeii})). A simple calculation yields the following types of
 ``strings''
 \begin{itemize}
 \item[{\bf (1)}] \underline{``Reals'':}

 unpaired, purely real spectral parameters $u_j$ and
 $\gamma_\beta$. 
 \item[{\bf (2)}] \underline{``Wide strings'':}

 ``type-I'': composites containing $n-1$ $\gamma$'s and $n$ $u$'s ($n>1$)
 \begin{eqnarray}
   u_{\alpha,k}^{(n,n-1)} &=& u_\alpha^{(n,n-1)} +i \left(n+1-2k\right),\quad
	   k=1\ldots n\nonumber\\
	   \gamma_{\alpha,j}^{(n,n-1)} &=& u_\alpha^{(n,n-1)} + i
	   \left(n-2j\right),\quad
	   j=1\ldots n-1\ ,\quad u_\alpha^{(n,n-1)}\in {\rm I\!R}
 \label{strings1}
 \end{eqnarray}

 ``type-II'': composites containing $n$ $\gamma$'s and $n-1$ $u$'s ($n>1$)
 \begin{eqnarray}
   \gamma_{\alpha,k}^{(n,n-1)} &=&  u_\alpha^{(n,n-1)} 
	 +i \left(n+1-2k\right),\quad
	   k=1\ldots n\nonumber\\
	   u_{\alpha,j}^{(n,n-1)} &=&  u_\alpha^{(n,n-1)} + i
	   \left(n-2j\right),\quad
	   j=1\ldots n-1\ ,\quad u_\alpha^{(n,n-1)}\in {\rm I\!R}
 \label{strings2}
 \end{eqnarray}

 \begin{figure}[ht]
 \begin{center}
 \noindent
 \includegraphics[scale=.5]{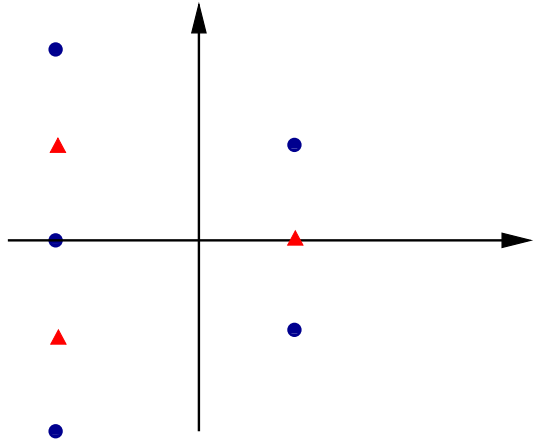}
 \end{center}
 \caption{\label{fig:widestr}
 Wide strings of lengths three and two respectively. The
 circles/triangles denote the positions of the $u$'s/$\gamma$'s
 involved in the string.
 }
 \end{figure}

 \item[{\bf (3)}] \underline{``Strange strings'':}

 composites containing $n$ $\gamma$'s and $n$ $u$'s ($n\geq 1$)
 \begin{eqnarray}
   u_{\alpha,k}^{(n,+)}&=& u_\alpha^{(n)} + i(n+1-2k-\frac{1}{2}),\quad
   k=1,\ldots,n\ ,\nonumber\\
   \gamma_{\alpha,k}^{(n,+)}&=& u_\alpha^{(n)} - i(n+1-2k-\frac{1}{2})
 \ ,\quad u_\alpha\in {\rm I\!R},
 \label{strings6}
 \end{eqnarray}
 or
 \begin{eqnarray}
   u_{\alpha,k}^{(n,-)}&=& u_\alpha^{(n)} - i(n+1-2k-\frac{1}{2}),\quad
   k=1,\ldots,n\ ,\nonumber\\
   \gamma_{\alpha,k}^{(n,-)}&=& u_\alpha^{(n)} + i(n+1-2k-\frac{1}{2})
 \ ,\quad u_\alpha\in {\rm I\!R},
 \label{strings7}
 \end{eqnarray}
 It is fundamental to observe that 
 solutions of this type are fundamentally  different
 from the  usual string solutions in that the set of roots on one level
 of the Bethe equations is {\sl not invariant under complex conjugation}. 
 The other solutions discussed above (reals and wide strings) are
 invariant under this operation. This is  similar to what was
 found recently for the anisotropic $sl(3)$ chain related with the 
 complex $SU(3)$ Toda theory in the continuum limit \cite{SW}. Of 
 course this 
 non invariance reflects the non hermitian nature of the superalgebra 
 hamiltonian: it casts doubts on analyzing the `thermodynamics' of 
 the system, for instance, but we shall see that naive calculations 
 seem to give the correct results anyway. 
 Note that although
 ``strange strings'' are not invariant under complex conjugation, the
 corresponding energy (\ref{energy})is still
 real because it depends symmetrically on the $u$ and $\gamma$ 
 parameters. 

 \begin{figure}[ht]
\includegraphics[scale=.5]{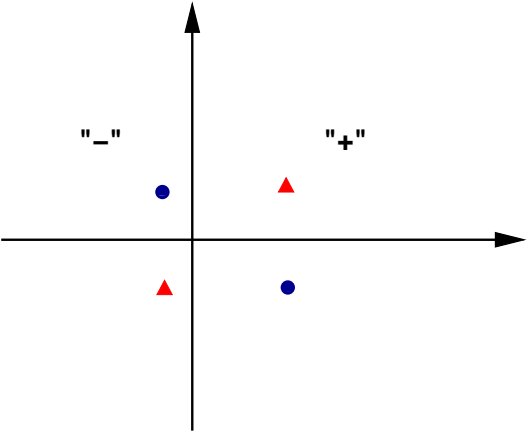}
  \caption{\label{fig:sstr2}
The two types of $n=1$ strange strings}
 \end{figure}
 
  \begin{figure}[ht]
\includegraphics[scale=.5]{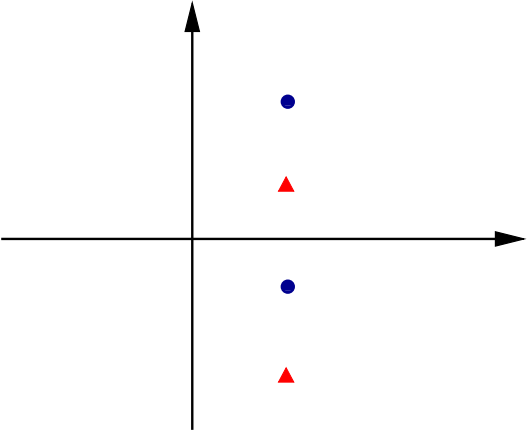}
  \caption{\label{fig:sstr}
A strange string with $n=2$.}
 \end{figure}

 \item[{\bf (4)}] \underline{``Narrow strings'':}

 composites containing $n$ $\gamma$'s and $n$ $u$'s ($n>1$)
 \begin{eqnarray}
   u_{\alpha,k}^{(n,n)} &=&  u_\alpha^{(n,n)} 
	 + \frac{i}{2}\left(n+1-2k\right),\quad
	   k=1\ldots n\nonumber\\
	   \gamma_{\alpha,j}^{(n,n)} &=&  u_\alpha^{(n,n)} +
	   \frac{i}{2}\left(n+1-2j\right),\quad
	   j=1\ldots n\ ,\quad u_\alpha^{(n,n)}\in {\rm I\!R}
 \label{strings3}
 \end{eqnarray}
 Narrow strings may be thought of as special cases of strange strings
 or wide strings in the following sense. 
 \begin{itemize}
 \item{}
 Combining a ``+''-strange string of length $n$ and centre $u^{(n)}$
 with a a ``-''-strange string of length $n$ and centre $u^{(n)}$ we
 obtain a narrow string of even length $2n$. This is shown for $n=1$
 in Fig.\ref{fig:narrowstr}(a). 
 \item{} 
 Combining a type-1 wide string of length n and centre $u^{(n,n-1)}$
 with a type-2 wide string of length n and centre $u^{(n,n-1)}$
 we obtain a narrow string of length $2n-1$. This is shown for the
 case $n=2$ in Fig. \ref{fig:narrowstr1}(b). 
 \end{itemize}

 \begin{figure}[ht]
 \begin{center}
\includegraphics[scale=.5]{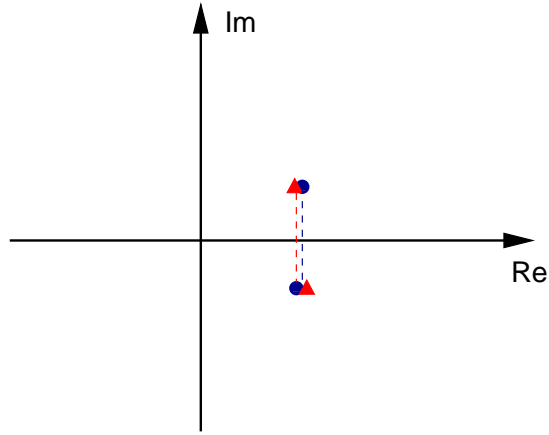}
 \end{center}
 \caption{\label{fig:narrowstr}
Combining a pair of ``+'' and ``-'' strange strings of length $1$
 gives a narrow string of length $2$;
 }
 \end{figure}
 
 \begin{figure}[ht]
 \begin{center}
\includegraphics[scale=.5]{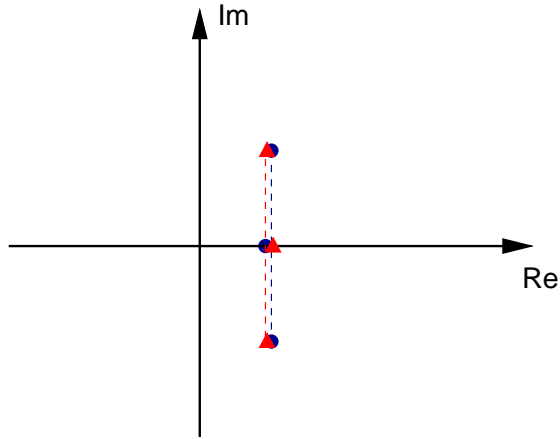}
 \end{center}
 \caption{\label{fig:narrowstr1}
 Combining a type-1 wide string on length 2 with a type-2 wide
 string of length 2 gives a narrow string of length $3$;
 }
 \end{figure}
 
 It is clear from our discussion that narrow strings are not
 ``fundamental'' but are merely degenerate cases of strange string
 solutions of the Bethe ansatz equations.
 \end{itemize}

 \section{TBA for the $3,\bar{3}$ and the $4$  (fundamental of OSp(2$|$2)) spin chains}
 
 We now derive  the TBA equations for the gapless spin chains to check (some aspects of) the completeness of our classification of solutions, and prepare subsequent discussion.

 To start we  derive equations for the densities of these various 
 solutions in the limit of large systems. This is done as usual by 
 considering that the  various roots in the complexes are slightly 
 off, and eliminating the corresponding small shifts through the Bethe 
 equations, resulting in `higher' Bethe equations. After taking the 
 logarithm and naively differentiating, we get a system which we write 
 symbolically as
 \begin{equation}
     \rho_{i}+\rho_{i}^{h}=s_{i}+\sum_{j}\Phi_{ij}\star\rho_{j}\label{symbol}
     \end{equation}
 (note we have included a $2\pi$ into the definition of the source and kernels). We introduce the fundamental objects
 \begin{equation}
     \alpha_{t}(u)={u+it/2\over u-it/2}
     \end{equation}
and the Fourier transformation
\begin{equation}
    \hat{f}(x)=\int du e^{iu x}f(u),~~~f(u)={1\over 2\pi}\int dx 
    e^{-iux}f(x)
    \end{equation}
We also set
\begin{equation}
a_{t}(u)={i\over 2\pi}{d\over du}\ln[\alpha_{t}(u)]={1\over 2\pi}{t\over 
u^{2}+t^{2}/4}
\end{equation}
so that $\hat{a}_{t}(x)=e^{-t|x|/2}$. Let us now give explicitly the terms in (\ref{symbol}).

The source terms (that is, the terms in the left hand side of the 
Bethe equations) in the case of alternating $3$ and $\bar{3}$ (\ref{baei}) for $n$ wide strings of type I or type II are of the 
form $\alpha_{2n-2}\alpha_{2n}$ which gives, after taking the log and 
differentiating, source terms of the form
\begin{equation}
    \hat{s}_{I,n}^{w}=\hat{s}_{II,n}^{w}=a_{2n}+a_{2n-2},~~~n\geq 1
    \end{equation}
with the convention that $a_{0}=0$, and that wide strings with $n=1$ 
are just real strings. Similarly, the source terms for $n$ strange 
strings of type I or type II are still complex numbers of modulus 
one, $e_{2n-1}e_{2n+1}$, and give after taking the logarithm and 
differentiating, real source terms  of the form
\begin{equation}
    \hat{s}_{I,n}^{s}=\hat{s}_{II,n}^{s}=a_{2n-1}+a_{2n+1},~~~n\geq 1
    \end{equation}
The kernel of interaction between $n$ wide strings of type I and $p$ 
wide strings of type I reads
\begin{equation}
    \alpha_{2p-2n}\alpha^{2}_{2p-2n+4}\ldots \alpha^{2}_{2p+2n-4}
    \end{equation}
while the interaction between $n$ wide strings of type I and $p$ wide 
strings of type II is instead
\begin{equation}
    \alpha_{2p+2n-2}\alpha^{2}_{2p-2n-6}\ldots \alpha^{2}_{2p-2n+2}
       \end{equation}
and there is a symmetry between type I and type II strings. 

Taking the logarithm and differentiating  gives the Fourier transforms 
of the functions $\Phi$ in (see \ref{symbol})
\begin{eqnarray}
    \hat{\Phi}_{I,n;I,p}^{w;w}=\hat{\Phi}_{I,p;I,n}^{w;w}=-2 {\sinh 
    nx\over\sinh x}e^{-(p-1)|x|}+e^{-(p-n)|x|}+\delta_{np},~~~p\geq 
    n\nonumber\\
    \hat{\Phi}_{I,n;II,p}^{w,w}=\hat{\Phi}_{II,p;I,n}^{w,w}=-2{\sinh 
    nx\over\sinh x}e^{-|p|x}-e^{-(p+n-1)|x|},~~~p\geq n
    \end{eqnarray}
Note that 
\begin{equation}
    \hat{\Phi}_{I,n;I,p}^{w;w}+\hat{\Phi}_{I,n;II,p}^{w,w}=-A_{2n-1,2p-1}+\delta_{n,p}
    \end{equation}
where
\begin{equation}
    A_{r,s}\equiv 2{\cosh x/2\over\sinh 
    x/2} \sinh[rx/2]e^{-s|x|/2},~~~s\geq r,~~~A_{r,s}=A_{s,r}
    \end{equation}
and $-A_{rs}+\delta_{rs}$ is  the Fourier transform of the 
standard kernel  $\Phi_{rs}$ describing interaction between $r$ and 
$s$ strings  in  the 
standard Bethe ansatz for XXX chain. Of course, this result should not 
be a surprise: concatenating $u$ and $\gamma$ variables transforms a 
$n$ wide string of type I or II into a $2n-1$ string for the XXX 
equations. Meanwhile, the source terms also match, for the chain acts 
on spin one representations.

We now turn to strange strings. Some of the kernels then will not be real, but 
after concatenation of the solutions, they will correspond to strings 
with even labels in the XXX model, a feature already seen for the 
source terms. For instance the interaction of type I $n$ strange 
strings with type I $p$ wide strings corresponds to the kernel
\begin{equation}
    {[u+i(p-n-{1\over 2})][u+i(p-n+{3\over 2})]^{2}\ldots
    [u+i(p+n-{5\over 2})]^{2}[u+i(p+n-{1\over 2})]\over
    [u-i(p+n-{3\over 2})]^{2}\ldots[u-i(p-n+{1\over 2})]^{2}}
    \end{equation}
    Meanwhile the interaction of type I $n$ strange strings with type II 
    $p$ wide strings is exactly the {\sl inverse of the complex 
    conjugate} of this quantity.
    After taking  the usual logarithm and differentiating one finds 
    the  
Fourier transforms of the functions $\Phi$ (\ref{symbol})
\begin{eqnarray}
    \hat{\Phi}_{I,n;I,p}^{s;w}={1\over 2}\left[-A_{2n,2p-1}-\hbox{sign 
    }(x) B_{2n,2p-1}\right]\nonumber\\
    \hat{\Phi}_{I,n;II,p}^{s;w}={1\over 2}\left[-A_{2n,2p-1}+\hbox{sign 
	}(x) B_{2n,2p-1}\right]\label{kersig}
    \end{eqnarray}
where 
\begin{equation}
    B_{r,s}={2\sinh x/2\over\cosh x/2}\sinh 
    (rx/2)e^{-s|x|/2},~~~s\geq r>1,~~~B_{r,s}=B_{s,r}
    \end{equation}
Observe the sum rule
\begin{equation}
    \hat{\Phi}_{I,n;I,p}^{s;w}+\hat{\Phi}_{I,n;II,p}^{s;w}=-A_{2n,2p-1}
    \end{equation}
   The interaction with real strings is a bit different, as 
\begin{equation}
    B_{1,s}=2\sinh{|x|\over 2}  e^{-s|x|/2}
    \end{equation}
Finally, we need the interaction of strange strings among 
themshelves. One finds
\begin{eqnarray}
    \hat{\Phi}_{I,n;I,p}^{s;s}=-C_{n,p}+\delta_{n,p}\nonumber\\
    \hat{\Phi}_{I,n;II,p}^{s;s}=-D_{n,p}
    \end{eqnarray}
with
\begin{eqnarray}
    C_{p,n}=
    {2\cosh x\over\sinh x} \sinh nx 
    e^{-p|x|},~~~p\geq n,~~~C_{p,n}=C_{n,p}\nonumber\\
    D_{p,n}= {2\over\sinh x} \sinh nx 
       e^{-p|x|},~~~p\geq n,~~~D_{p,n}=D_{n,p}
       \end{eqnarray}
Note that, once again,
\begin{equation}
    \hat{\Phi}_{I,n;I,p}^{s;s}+\hat{\Phi}_{I,n;II,p}^{s;s}=\delta_{np}-C_{n,p}-D_{n,p}=
    \delta_{np}-A_{2n,2p}
    \end{equation}

       We now write symbolically the system of equations as in 
       (\ref{symbol}) but allowing explicitely for the two types of 
       solutions:
\begin{eqnarray}
    \rho_{I,i}+\rho_{I,i}^{h}=s_{I,i}+\sum_{j}\Phi_{I,i;I,j}\star\rho_{I,j}+
    \Phi_{I,i;II,j}\star\rho_{II,j}\nonumber\\
    \rho_{II,i}+\rho_{II,i}^{h}=s_{II,i}+\sum_{j}\Phi_{II,i;I,j}\star\rho_{I,j}+
       \Phi_{II,i;II,j}\star\rho_{II,j}\label{denseq}
       \end{eqnarray}
and the labels $i,j$ are short hands for $n,p$ and $s,w$. The energy 
is given by (we often do not write the integration variable)
\begin{equation}
    {E\over L}=\int \sum_{i} e_{I,i}~\rho_{I,i}+e_{II,i~}\rho_{II,i}\label{energy1}
    \end{equation}
    where $e_{I,n}^{w}=e_{II,n}^{w}=-2 
    s_{I,n}^{w}=-2s_{II,n}^{w}$, and similarly for the 
    strange strings. Note that although the Bethe solutions for 
    strange strings are not invariant under complex conjugation, the 
    symmetry between the two types of roots in the expression of the 
    energy guarantees that it  is {\sl real}. Note we have adjusted the sound velocity (using the results of \cite{EFS} for low energy excitations) so that the theory is isotropic in the continuum limit.
    
    We must now pause to discuss the effects of the strange strings more carefully. Since some of the corresponding bare scatterings are not actually pure phases, the associated kernels are not real (this translates into some of the Fourier transforms having an even and an odd part under change of sign of the variable $x$; see (\ref{kersig}) for instance). On the other hand, the densities are defined as variations of integers with respect to a real variable (the center of the strings), and thus are real. The equations (\ref{denseq}) can only make sense if the right hand sides are real as well.  It seems reasonable to solve this constraint by demanding that the densities of  type $I$ and type $II$   $n$ strange strings  are equal (as functions of the rapidities). 
    
The densities being real, the entropy has the usual expression in the TBA:
\begin{equation}
{S\over L}=\int \sum_i (\rho_{I,i}+\rho_{I,i}^h)\ln(\rho_{I,i}+\rho_{I,i}^h)-\rho_{I,i}\ln\rho_{I,i}-\rho_{I,i}^h\ln\rho_{I,j}^h+I \leftrightarrow II\label{entropy}
\end{equation}
Note here that one has to be careful  to not constrain the individual type $I$ and type $II$ strings to have the same centers, as this 
would reduce the entropy by a factor two, and give rise to the wrong result in the end. Rather, we use the  expressions (\ref{denseq},\ref{energy}, \ref{entropy}) and minimize the free energy $F=U-TS$ as a function of the densities $\rho_{I,i},\rho_{II,i}$ subject to the constraint that the densities of strange strings be equal. This is in fact a transparent subtlety, if  the naive extremum of $F$ without any constraint turns out to be symmetric under $I\leftrightarrow II$.

Introducing pseudoenergies through 
$\rho_{n}^{h}/\rho_{n}=e^{\epsilon_{n}/T}$, we find the (formal) TBA equations
\begin{eqnarray}
    \epsilon_{I,i}=e_{I,i}-
    T\sum_{j}\Phi_{I,i;I,j}\star\ln\left(1+e^{-\epsilon_{I,j}/T}\right)
    -  T\sum_{j}\Phi_{II,i;I,j}\star\ln\left(1+e^{-\epsilon_{II,j}/T}\right)\nonumber\\
    \epsilon_{II,i}=e_{II,i}-
       T\sum_{j}\Phi_{I,i;II,j}\star\ln\left(1+e^{-\epsilon_{I,j}/T}\right)
       -  
       T\sum_{j}\Phi_{II,i;II,j}\star\ln\left(1+e^{-\epsilon_{II,j}/T}\right)\label{tbai}
        \end{eqnarray}
while 
\begin{equation}
    {F\over L}=-T\sum_{i}\int\left[s_{I,i}\ln\left(1+e^{-\epsilon_{I,i}/T}\right)
    +s_{II,i}\ln\left(1+e^{-\epsilon_{II,i}/T}\right)\right]
    \end{equation}
Demanding   $\epsilon_{I,i}=\epsilon_{II,i}\equiv \epsilon_{i}$ at 
equilibrium, we get a new system which reads
\begin{equation}
    \epsilon_{i}=e_{i}-\sum_{j}
    T(\Phi_{I,i;I,j}+\Phi_{II,i;I,j})\star\ln\left(1+e^{-\epsilon_{j}/T}\right)
    \end{equation}
    together with
\begin{equation}
    {F\over L}=-2T\sum_{i}\int s_{i}\ln\left(1+e^{-\epsilon_{i}/T}\right)
    \end{equation}
 Using the sum rules mentioned previously, it is clear that 
 the resulting TBA equations {\sl coincide with those for the spin 
 one XXX chain}, while  the free energy is {\sl twice} the free energy of 
 that chain, leading to a central charge $c=2\times {3\over 2}=3$ in 
 the UV. 
 
\begin{figure}
 \centering
     \includegraphics[scale=.5]{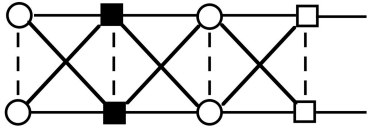}
 \caption{TBA diagram for the $sl(2/1)$ chain with alternating $3$ 
 and $\bar{3}$ representations. }
 \label{tbadiag1}
 \end{figure}

 Let us be more explicit on these points. By simple manipulations 
 ones finds the universal form of (\ref{denseq}) 
 \begin{eqnarray}
     2\cosh {x\over 
     2}\left(\widehat{\rho}_{I,i}+\widehat{\rho}_{II,i}\right)=2\delta_{i,2}+\widehat{\rho}_{I,i-1}^{h}+
    \widehat{\rho}_{I,i+1}^{h}
    \widehat{\rho}_{II,i-1}^{h}+
    \widehat{ \rho}_{II,i+1}^{h}
     \end{eqnarray}
 and similarly the universal form of (\ref{tbai}) 
 \begin{eqnarray}
    \epsilon_{I,i}+\epsilon_{II,i}=
     -{4s\over\pi}\delta_{i,2}+Ts\star\left[\ln\left(1+e^{\epsilon_{I,i-1}/T}\right)
     +\ln\left(1+e^{\epsilon_{I,i+1}/T}\right)\right.\nonumber\\
   \left.+ \ln\left(1+e^{\epsilon_{II,i-1}/T}\right)
	  +\ln\left(1+e^{\epsilon_{II,i+1}/T}\right)\right]
	  \end{eqnarray}
with $\widehat{s}={1\over 2\cosh x/2}$. It is convenient to represent these equations by a TBA diagram as 
shown on the figure \ref{tbadiag1}. Dots stand for wide strings and 
squares for strange strings. Note how they nicely get organized into a 
pattern exactly identical to the one of the XXX chain. At $T=0$ the ground state is 
obtained by filling up states corresponding to the black squares. Of 
course the presence of two lines of nodes corresponds to the algebra 
being of rank two; moreover the symmetry between the two lines arises 
from our symmetric choice of roots. 

 It is a simple matter to extend these results to a chain with 
 alternating fully symmetric and antisymmetric representations, 
 represented by  Young diagrams with $p$ boxes . The TBA in universal form is 
 represented by diagrams identical to the foregoing ones, only the 
 massive nodes are the $2p^{th}$ ones.   The central 
 charge in the UV is $c={6p\over p+2}$.
 
\begin{figure}
  \centering
      \includegraphics[scale=.5]{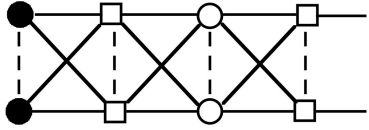}
  \caption{TBA diagram for the $OSP(2/2)$  chain (four dimensional 
  fundamental representation). }
  \label{tbadiag}
  \end{figure}

 It is also easy to extend the results to the case of the fundamental representation. 
The 
 TBA now has source terms on the first nodes, as represented on 
 figure \ref{tbadiag}.  The UV central charge is $c=2$.

 The TBA for the spin chain confirms identifications of central 
 charges carried out by other means in \cite{EFS},\cite{JS}. Moreover,
 it constitutes a crucial check of completeness of the solutions of 
 the Bethe ansatz equations, and we can now use our strings 
 classification to tackle the more interesting problem of scattering 
 in $osp(2/2)$ integrable quantum field theories.

Nevertheless, it is fair to remind the reader that the TBA is rather 
 insensitive to details of the diagonalization. Consider for instance 
 the well known $SU(2)$ spin 1/2 XXX chain. The total number of 
 states with $S^{z}=0$ is  $\Omega_{S^{z}=0}=\left(\begin{array}{c}
 L/2 \\
 L\end{array}\right)$ while the number of highest weight states with 
 $j=0$ is $\Omega_{j=0}=\left(\begin{array}{c}
 L/2 \\
 L\end{array}\right)-=\left(\begin{array}{c}
 L/2-1 \\
 L\end{array}\right)$. However, if we consider the entropies 
 associated with these two numbers we have 
 \begin{equation}
     \ln \Omega_{j=0}=\ln\Omega_{S^{z}=0}-\ln{L+2\over 4}
     \end{equation}
 so the difference is of order ${\ln L\over L}$ and goes to zero as 
 $L\rightarrow\infty$. In other words, thermodynamics based on a 
 scheme that would not provide all eigenstates but only the highest 
 weight states say would still reproduce the expected results. 
In particular, even if our string analysis turns out to be wrong or 
 incomplete for states in indecomposable representations, it is still 
 very conceivable - as we indeed observe - that this effect is 
 negligible in the thermodynamic limit. 
 
 A related comment concerns the reality condition we imposed on the densities of the two types of strings. Since evidence is strong that the minimum of the free energy functional is symmetric under the exchange of the two types of strings, our condition led us to what should be the correct result. It is not clear however how to describe solutions where there is a dissymmetry between type I and type II strange strings. Presumably, the string hypothesis breaks down in this case.

   \section{TBA for OSp(2$|$2) scattering: the fundamental 
   representation case and the GN model}

   We now consider the different but related problem of writing up the TBA for the $OSp(2|2)$ GN model. The  action is 
\begin{equation}
S=\int {d^2x\over 2\pi}  \left[\sum_{i=1}^2\psi^i_L\partial\psi^i_L+
\psi^i_R\bar{\partial}\psi^i_R+\beta_L\partial\gamma_L+\beta_R\bar{\partial}\gamma_R
+g
\left(\psi^1_L\psi^2_R-\psi^2_L\psi^1_R+\gamma_L\beta_R-\beta_L\gamma_R\right)^2\right]
\end{equation}
where the $\psi$ are Majorana fermions of conformal weight $1/2$, and
the $\beta\gamma$ are bosonic ghosts of weight $1/2$ as
well.  The theory at $g=0$ can be identified with the $OSp(2|2)$ WZW model at level $k=1$ (part of the literature denotes this  level $k=-1/2$ instead, focussing on the sub $SU(2)$). The lowest order beta function is $\beta(g)\propto -g^2$ and for $g<0$ the theory is massive \footnote{We use Majorana instead of complex fermions. Some of the signs are therefore switched when compared to  \cite{BassiLeclair}.}.
The corresponding S matrix has been proposed  by  Bassi  and Leclair in \cite{BassiLeclair}. It involves, as is usual for  $O(N)$ GN models at large $N$ but not, a priori, at small $N$, particles of mass $M$  in the four dimensional fundamental representation, with scattering matrix of the form 
   $S(\theta)=Z(\theta)R_{\hbox{fund}}(\theta)$ where $Z$ is 
a normalization factor, and $R_{\hbox{fund}}$ is the solution of the Yang Baxter 
equation for the fundamental representation, normalized such that the scattering kernel for particles in 
the highest weight state $b=0,S^{z}=1/2$ is equal to $Z$ (that is the 
matrix element in $R$ is equal to unity). The $R$ matrix is the same 
as the matrix defining the Boltzmann weights of the integrable spin 
chain. in the fundamental representation. The minimal $Z$ factor (in general we denote by $Z$ factor the element of the S matrix corresponding to diagonal scattering of two particles of the same type - no ambiguity will arise in what follows) given in \cite{BassiLeclair} using standard crossing and unitarity arguments turns out to be the {\sl square} of the similar factor for the $SU(2)_1$ WZW theory, or sine-Gordon model at $\beta^2=8\pi$,
\begin{equation}
    Z=\left[{\Gamma\left(1+{i\theta\over 
    2\pi}\right)\Gamma\left({1\over 2}-{i\theta\over 2\pi}\right)\over
    \Gamma\left(1-{i\theta\over 2\pi}\right)\Gamma\left({1\over 
    2}+{i\theta\over 2\pi}\right)}\right]^{2}
    \end{equation}
Doubts were raised in \cite{Birgit1} about this S matrix, on the grounds that, for $OSp(2|2)$ as for the usual $O(N)$ GN models at small $N$, the fundamental particles might well be kinks, while the fundamental fermions (and bosons in the super case) are unstable. It was suggested that maybe the S matrix of   \cite{BassiLeclair} in fact describes the sigma model  $OSp(2|2)/OSp(1/2)$ instead. We will, in what follows, find evidence that this expectation is not correct, and that the S matrix of \cite{BassiLeclair}
does indeed describe the GN model. 
Of course, writing the TBA for the scattering theory is not the same as writing it for the integrable spin chain, as  the number of particles is allowed to fluctuate.  The two problems are however closely related, as is well known (see \cite{PaulKen} for a pedagogical discussion)

The problem of quantizing the particles turns, in the 
continuous limit, into the basic equation
\begin{equation}
    2\pi(\sigma+\sigma^{h})=M\cosh\theta+{1\over i}{d\over d\theta}\ln 
    Z\star \sigma+{1\over i}{d\over d\theta}\ln \Lambda
    \end{equation}
Here $\Lambda$ is the eigenvalue of the  monodromy matrix, ie of the 
matrix describing the process of passing one particle through a 
system with a certain number of particles already. This eigenvalue (properly normalized, so that it is equal to unity if all particles are of the same type, corresponding to `all spins up' in the auxiliary problem) is 
found itself by another, `auxiliary' Bethe ansatz, which coincides 
with the one used to diagonalize the spin chain. We need the general formula 
for the eigenvalue picked up when passing a particle of rapidity 
$v$ through a set of particles with rapidities $\{v_{A}\}$ 
\begin{equation}
    \Lambda(v|\{v_{A}\})=\prod {v-u_{j}+i/2\over 
    v-u_{j}-i/2}{v-\gamma_{\alpha}+i/2\over v-\gamma_{\alpha}-i/2}
    \end{equation}
 (one checks that its logarithmic derivative gives 
the energy for the hamiltonian in the previous section), where the 
$u,\gamma$ are related to the $\{v_{A}\}$  through the Bethe equations 
\begin{eqnarray}
    \prod_{A}{u_{j}-v_{A}+i/2\over 
    u_{j}-v_{A}-i/2}=\prod_{\beta=1}^{M}{u_{j}-\gamma_{\beta}+i\over
    u_{j}-\gamma_{\beta}-i},~~~j=1,\ldots,N\nonumber\\
    \prod {\gamma_{\alpha}-v_{A}+i/2\over 
       \gamma_{\alpha}-v_{A}-i/2}=\prod_{k=1}^{N}{\gamma_{\alpha}-u_{k}+i\over
       \gamma_{\alpha}-u_{k}-i},~~~\alpha=1,\ldots,M\label{auxiliary}
   \end{eqnarray}
Note that the {\sl same set} of rapidities $\{v_{A}\}$ is used in 
both equations for the $u$ as well as the $\gamma$'s. 
Here we have used the lattice rapidities for the auxiliary problem to match with the previous section. They 
 are related to the 
relativistic ones by $\theta=\pi u$. We define the Fourier 
transformation for functions of $\theta$ as
\begin{equation}
    \hat{f}(x)=\int d\theta e^{i \theta x/\pi}f(\theta)
    \end{equation}
The Bethe equations for the auxiliary system (\ref{auxiliary}) admit the same 
clasification of roots as the one for the lattice model, and lead therefore 
to the following system
\begin{eqnarray}
    \rho_{I,n}^{w}+\hat{\rho}_{I,n}^{w}&=&a_{2n-1}\star\sigma 
    +\sum_{p} \Phi_{I,n;I,p}^{w;w}\star\rho_{I,p}^{w}+
    \Phi_{I,n;II,p}^{w;w}\star\rho_{II,p}^{w}\nonumber\\&+&
    \Phi_{I,n;I,p}^{w;s}\star\rho_{I,p}^{s}+
    \Phi_{I,n;I,p}^{w;s}\star\rho_{II,p}^{s}\nonumber\\
    \rho_{I,n}^{s}+\hat{\rho}_{I,n}^{s}&=&a_{2n}\star\sigma 
	+\sum_{p} \Phi_{I,n;I,p}^{s;w}\star\rho_{I,p}^{w}+
	\Phi_{I,n;II,p}^{s;w}\star\rho_{II,p}^{w}\nonumber\\&+&
	\Phi_{I,n;I,p}^{s;s}\star\rho_{I,p}^{s}+
	\Phi_{I,n;I,p}^{s;s}\star\rho_{II,p}^{s}\label{good}
	\end{eqnarray}
	and a similar equation for type II roots  (the kernels $\Phi$ have been defined in the previous sections). As for the $\sigma$ 
	equations they are \footnote{The mass parameter $M$ should not be confused with the number of roots of type $\gamma$.}
 \begin{eqnarray}
     \sigma+\sigma^{h}={M\over 2\pi}\cosh\theta+{1\over 2i\pi}{d\over d\theta}\ln 
     Z\star\sigma -\sum_{k\geq 
     1}a_{2k-1}\star(\rho_{I,k}^{w}+\rho_{II,k}^{w})\nonumber\\
     -\sum_{k\geq 1}
     a_{2k}\star(\rho_{I,k}^{s}+\rho_{II,k}^{s})\label{notquite}
\end{eqnarray}
where we now define
\begin{equation}
a_{t}(\theta)={i\over 2\pi}{d\over d\theta}\ln {\theta+i\pi t/2\over \theta-i\pi t/2}
\end{equation}
and all the densities are now in terms of the (physical or auxiliary) rapidities.

 We write symbolically our system 
as
\begin{eqnarray}
    \rho_{I,i}+\rho_{I,i}^{h}=\Phi_{I,i;\sigma}\star\sigma+\sum_{j}\Phi_{I,i;I,j}\star\rho_{I,j}+
    \Phi_{I,i;II,j}\star\rho_{II,j}\nonumber\\
    \rho_{II,i}+\rho_{II,i}^{h}=\Phi_{II,i;\sigma}\star\sigma+\sum_{j}\Phi_{II,i;I,j}\star\rho_{I,j}+
\Phi_{II,i;II,j}\star\rho_{II,j}\nonumber\\
\sigma+\sigma^{h}={M\cosh\theta\over 2\pi}+\Phi_{\sigma,\sigma}\star\sigma+\sum_{i}\Phi_{\sigma;I,i}\star\rho_{I,i}
+\Phi_{\sigma;II,i}\star\rho_{II,i}
\end{eqnarray}
and the labels $i,j$ are short hands for $n,p$ and $s,w$. We analyze then 
the thermodynamics of this system, with the energy 
\begin{equation}
    {E\over L}=\int \sigma(\theta) M\cosh\theta d\theta
    \end{equation}
Introducing pseudoenergies through 
$\rho^{h}/\rho=e^{\epsilon_{n}/T}$, 
$\sigma^{h}/\sigma=e^{\epsilon_{\sigma}/T}$, we find the TBA equations
\begin{eqnarray}
    \epsilon&=&M\cosh\theta-T\Phi_{\sigma,\sigma}\star\ln\left(1+e^{-\epsilon_{\sigma}/T}\right)
    -T\sum_{i}\Phi_{I,i;\sigma}\star\ln\left(1+e^{-\epsilon_{I,i}/T}\right)\nonumber\\
    &-&T\Phi_{II,i;\sigma}\star\ln\left(1+e^{-\epsilon_{II,i}/T}\right)\nonumber\\
    \epsilon_{I,i}&=&-T\Phi_{\sigma;I,i}\star\ln\left(1+e^{-\epsilon_{\sigma}/T}\right)
    -T\sum_{j}\Phi_{I,j;I,i}\star\ln\left(1+e^{-\epsilon_{I,j}/T}\right)\nonumber\\
   &-&  T\sum_{j}\Phi_{II,j;I,i}\star\ln\left(1+e^{-\epsilon_{II,j}/T}\right)\nonumber\\
    \epsilon_{II,i}&=&-T\Phi_{\sigma;II,i}\star\ln\left(1+e^{-\epsilon_{\sigma}/T}\right)
    -
T\sum_{j}\Phi_{I,j;II,i}\star\ln\left(1+e^{-\epsilon_{I,j}/T}\right)\nonumber\\
&-&  T\sum_{j}\Phi_{II,j;II,i}\star\ln\left(1+e^{-\epsilon_{II,j}/T}\right)
 \end{eqnarray}
while 
\begin{equation}
    {F\over L}=-T\int {M\cosh\theta\over 
    2\pi}\ln\left(1+e^{-\epsilon_{\sigma}/T}\right)d\theta
    \end{equation}
Observing now that 
$\Phi_{\sigma;I,i}=\Phi_{\sigma;II,i}\equiv\Phi_{\sigma,i}$, we 
expect that $\epsilon_{I,i}=\epsilon_{II,i}\equiv \epsilon_{i}$ at 
equilibrium, and thus get a new system 

\begin{eqnarray}
    \epsilon_{i}=-\sum_{j}
    T(\Phi_{I,j;I,i}+\Phi_{II,j;I,i})\star\ln\left(1+e^{-\epsilon_{j}/T}\right)
    -T\Phi_{\sigma;i}\star\ln\left(1+e^{-\epsilon_{\sigma}/T}\right)\nonumber\\
    \epsilon_{\sigma}=M\cosh\theta-T\Phi_{\sigma,\sigma}\star
    \ln\left(1+e^{-\epsilon_{\sigma}/T}\right)
    -2T\sum_i
    \Phi_{i;\sigma}\star\ln\left(1+e^{-\epsilon_{i}/T}\right)
\end{eqnarray}
Remarkably, this TBA is identical with the TBA one would write for 
the $SU(2)$ PCM model  \cite{ZamoZamosigma}. There, the scattering matrix is the product of 
two $SU(2)$ isotropic sine-Gordon scattering matrices, and the 
strings involved in diagonalizing each of the two scattering matrices 
behave like  our type $I$ and type $II$ solutions.

\begin{figure}
 \centering
     \includegraphics[scale=.5]{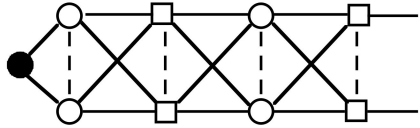}
 \caption{TBA diagram for the $OSP(2/2)$ 
 theory with particles in the fundamental representation. }
 \label{tbadiag2}
 \end{figure}
 
\begin{figure}
 \centering
     \includegraphics[scale=.5]{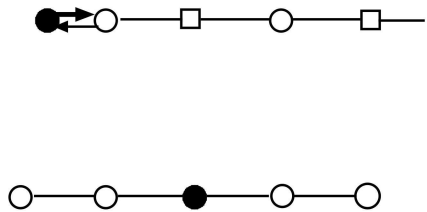}
 \caption{Equivalent TBA diagrams for the $OSP(2/2)$ 
 theory with particles in the fundamental representation, showing 
  mapping to the $SU(2)$ PCM model. }
 \label{tbadiag3}
 \end{figure}

Identifying the nodes of type I and type II and using the fact that 
the pseudoenergies are equal in equilibrium gives the equivalent 
diagrams shown in \ref{tbadiag3}.

We conclude that $c_{UV}=3$, for our scattering theory. This is in agreement with the effective central charge for the GN model in the sector with anti-periodic fermions (NS), and provides some justification for the S matrix of \cite{BassiLeclair}.

If we now accept that the S matrix in \cite{BassiLeclair} is the correct one, 
it follows  from our calculation  that 
the free energy of the $OSP(2/2)$ GN model in the NS sector  is exactly equal to the one 
of the $SU(2)$ PCM model. Now note the latter is the same as the $SO(4)/SO(3)$ sigma model: we thus have established part of the equivalences expected from the introduction.

Our observation deserves several remarks. The first is, that the torus partition function of the free GN model with periodic boundary conditions for the fermions is exactly equal to unity. This results from the cancellation between the fermionic and bosonic contributions (each equal to $\hbox{Det }\Delta^{\pm 1}$, $\Delta$ the Laplacian), and is in fact what motivates the appearance of the $OSp(2|2)$ symmetry in the SUSY approach to disordered systems.  One can easily show that this supersymmetry is not broken by the introduction of a current-current perturbation. The partition function with periodic BC should thus remain unity, which implies in particular that the {\sl bulk} term in the free  energy - which does not depend on the boundary conditions - should vanish exactly. Turning now to the non trivial sector with antiperiodic boundary conditions for the fermions, the free energy now becomes the non trivial object we have studied in this section, but its  bulk part {\sl should still vanish exactly}.  Since we have argued the free energy for the $OSp(2|2)$ GN model should coincide with the one of the $SU(2)$ PCM, this implies the bulk term must vanish there. To calculate this bulk term, we use the fact that $SU(2)$ PCM coincides with the $O(4)$ sigma model. The bulk term for $O(N)$ sigma models follows from the general results in 
  \cite{Alyosha},\cite{Fateevbulk},\cite{Hasenfratz}:  
  \begin{equation}
  {F^{(0)}\over L}={M^2\over 8}\cot {\pi\over N-2}\label{general}
  \end{equation}
  where $M$ is the lowest mass. We conclude that for the $SU(2)$ PCM, for which $N=4$:
\begin{equation}
{F^{(0)}\over L}={M^2\over 8}\times 0
\end{equation}
 in agreement with our expectation for $OSp(2|2)$.

\section{The OSp(2$|$2) supersphere sigma model TBA.}

If the S matrix proposed in \cite{BassiLeclair} is for the GN model indeed, the question is now open as to what the S matirx of the sigma model might be. By analogy with what happens for $O(N)$, $N>0$, and after switching the roles of GN and $\sigma$M, one would expect that the S matrix for the $OSp(2|2)$ sigma model should involve spinor representations. This could be the three dimensional representations (fundamental of $SL(2|1)$ and its conjugate), or maybe some infinite dimensional representations\cite{spinor}. Our naive attempts in this direction have not been successful however. 

An alternative approach consists in starting from  lattice regularizations.  It was indeed argued in \cite{Birgit1} that the continuum limit of the integrable {\sl staggered} $OSp(2|2)$ chain built on the fundamental was described by the $OSp(2|2)/OSp(1|2)$ sigma model in the massive regime. The staggering is similar to what has been studied in particular under the name of lattice light cone regularization for a large class of perturbed CFTs and sigma models \cite{ReshetikhinSaleur}\cite{Destri}. For a recent reference in a very similar case, see
\cite{Hergedus}.

Recall that, for the integrable chain based on the four dimensional representation,  the bare source term involves a factor 
${i\over 2}$ instead of $i$, and as a result, the source term for the 
wide and strange strings are different:
\begin{eqnarray}
    \hat{s}^{w}_{I,n}&=&\hat{s}^{w}_{II,n}=a_{2n-1}\nonumber\\
    \hat{s}^{s}_{I,n}&=&\hat{s}^{s}_{II,n}=a_{2n}
    \end{eqnarray}
(so in general we can write that the source term of a string with $p$ 
roots is $a_{p}$).

What we have to do, following the general ideas of \cite{Birgit1},\cite{Hergedus} is to consider instead of the homogeneous chain, the chain arising from staggering spectral parameters. The relevant equations now read
\begin{eqnarray}
    \left({u_{j}+\Lambda/2+i/2\over 
    u_{j}+\Lambda/2-i/2}\right)^{L/2}
     \left({u_{j}-\Lambda/2+i/2\over 
    u_{j}-\Lambda/2-i/2}\right)^{L/2}=\prod_{\beta=1}^{M}{u_{j}-\gamma_{\beta}+i\over
    u_{j}-\gamma_{\beta}-i},~~~j=1,\ldots,N\nonumber\\
    \left({\gamma_{\alpha}+\Lambda/2+i/2\over 
       \gamma_{\alpha}+\Lambda/2-i/2}\right)^{L/2}
        \left({\gamma_{\alpha}-\Lambda/2+i/2\over 
       \gamma_{\alpha}-\Lambda/2-i/2}\right)^{L/2}=\prod_{k=1}^{N}{\gamma_{\alpha}-u_{k}+i\over
       \gamma_{\alpha}-u_{k}-i},~~~\alpha=1,\ldots,M\label{bae}
   \end{eqnarray}
with the quantum numbers $B={N-M\over 
2}$and  $J^{3}=L-{M+N\over 2}$. The energy 
reads then
\begin{equation}
    E\propto -\left(\sum_{k=1}^{N}{1\over (u_{k}-\Lambda/2)^{2}+1/4}
    +{1\over (u_{k}+\Lambda/2)^{2}+1/4}
    +\sum_{\beta=1}^{M}{1\over 
    (\gamma_{\beta}-\Lambda/2)^{2}+1/4}+{1\over 
    (\gamma_{\beta}+\Lambda/2)^{2}+1/4}\right)\label{energy2}
\end{equation}
The ground state of the system was discussed in \cite{EFS,JS} and 
was shown to consist of filled up sea of real roots
(ie $n=1$ wide strings) of type I and type II. Excitations can easily be shown (see more details below) to be massive, with a mass scale $Ma\propto e^{-\pi\Lambda/2}$, with $a$ the lattice spacing, so the continuum limit is obtained as usual with $a\rightarrow 0$, $\Lambda\rightarrow\infty$. From the general discussion in the second section, it follows that the ground state energy is given by twice the ground state energy of the staggered XXX chain. But for the latter, the ground state energy is well known to reproduce, in the continuum limit where $\Lambda\rightarrow\infty$, the ground state energy of the $SU(2)_1$ WZW model perturbed by the current current interaction, or, equivalently, of the sine-Gordon model at $\beta_{SG}^2=8\pi$.  

It is also a well known fact that the $SO(4)$ GN model is equivalent to two decoupled sine Gordon models at their marginal point \cite{Shankar}. We thus have obtained our other claim, namely that the free energy of the $OSp(2|2)$ supersphere sigma model with antiperiodic boundary conditions coincides with the free energy of the $SO(4)$ GN model.

 We thus 
have coincidence  of free energies as
\begin{eqnarray}
    OSP(2/2)\hbox{ GN }\leftrightarrow  SO(4)/SO(3) \hbox{  sigma Model  }\equiv  SU(2)\times SU(2)\hbox{ PCM}\nonumber\\
    OSP(2/2)/OSP(1/2)\hbox{ supersphere sigma Model  }\leftrightarrow
    SO(4)\hbox{ GN}\equiv SU(2)\times SU(2)\hbox{ GN}
    \end{eqnarray}

 We note however that the relationship does not seem to extend to the 
 full spectra of the theories. We haven't been able to find an 
 $SU(2)\times SU(2)$ symmetry within either of the $OSP(2/2)$ models 
 for instance. The UV limits of the flows are quite different as 
 well. For instance the UV limit of the $SO(4)$ GN is compact, while the UV 
 limit of the $OSP(2/2)$ supersphere sigma model has a non compact 
 direction. 

\bigskip

Now we can get back to the question of what the S matrix might be. 
For ordinary models, this can be rather easily determined using the Bethe equations from the lattice regularization. The idea is to rewrite the continuous Bethe ansatz 
equations in terms of excitations over the vacuum \cite{Korepin}, ie by putting on 
the right hand side, not densities of $n=1$ wide strings, but 
densities of $n=1$ {\sl holes} in the wide strings ground state 
distribution, that is replace $\rho_{I,1}$ and $\rho_{II,1}$ by 
$\rho_{I,1}^{w,h}$ and $\rho_{II,1}^{w,h}$.

After laborious but straightforward calculations, one ends up with the 
following equations for the densities of real particles (we denote 
$\sigma_{I}\equiv \rho_{I,1}^{w}$ and similarly for type II)
\begin{eqnarray}
    \widehat{\sigma}_{I}+\widehat{\sigma}_{I}^{h}&=&{\cosh\Lambda x/2\over 2\cosh(x/2)}-{e^{-|x|}\over 
    2\sinh|x|}\widehat{\sigma}_{I}^{h}+{1\over 2\sinh |x|}\widehat{\sigma}_{II}^{h}-
    \sum_{k\geq 2}e^{-(k-1)|x|}\widehat{\rho}_{I,k}^{w}\nonumber\\
    &-&\sum_{k\geq 1} 
   e^{-(k-1/2)|x|}\left({1+\hbox{sign }(x)\over 2}\widehat{\rho}_{I,k}^{s}+
   {1-\hbox{sign }(x)\over 2}\widehat{\rho}_{II,k}^{s}\right)\nonumber\\
   \widehat{\sigma}_{II}+\widehat{\sigma}_{II}^{h}&=&{\cosh\Lambda x/2\over 2\cosh(x/2)}-{e^{-|x|}\over 
      2\sinh|x|}\widehat{\sigma}_{II}^{h}+{1\over 2\sinh 
      |x|}\widehat{\sigma}_{I}^{h}-
      \sum_{k\geq 2}e^{-(k-1)|x|}\widehat{\rho}_{II,k}^{w}\nonumber\\
      &-&\sum_{k\geq 1} 
     e^{-(k-1/2)|x|}\left({1+\hbox{sign }(x)\over 2}
     \widehat{\rho}_{II,k}^{s}+{1-\hbox{sign }(x)\over 
     2}\widehat{\rho}_{I,k}^{s}\right)\label{systemi}
\end{eqnarray}
For the other solutions, one observes that the bare source terms 
disappear, and are replaced by source terms which are convolutions 
involving the densities $\sigma_{I}^{h}$ and $\sigma_{II}^{h}$. As for 
the interactions between different types of roots, one finds that {\sl 
a shift of one unit in the length} is taking place. That is, what was 
before an $n$ ($n>1$) wide root of type I or II now behaves as an $n-1$ 
strange root of type I or II, while what was before a strange $n$  root of 
type I or II now behaves as an $n$ wide root of type I or II. In other 
words, we have  the following system
\begin{eqnarray}
    \widehat{\rho}_{I,k}^{w}+\widehat{\rho}_{I,k}^{w,h}&=&
    e^{-(k-1)|x|}\widehat{\sigma}_{I}^{h}+\sum_{l\geq 
    2}D\widehat{\Phi}^{w,w}_{I,k;I,l}\star\widehat{\rho}_{I,l}^{w}+\ldots\nonumber\\
    \widehat{\rho}_{II,k}^{w}+\widehat{\rho}_{II,k}^{w,h}&=&
       e^{-(k-1)|x|}\widehat{\sigma}_{II}^{h}+\sum_{l\geq 
       2}D\widehat{\Phi}^{w,w}_{II,k;I,l}\star\widehat{\rho}_{I,l}^{w}+\ldots\nonumber\\
       \widehat{\rho}_{I,k}^{s}+\widehat{\rho}_{I,k}^{s,h}&=&e^{-(k-1/2)|x|}
       \left({1+\hbox{sign}(x)\over 
       2}\widehat{\sigma}_{I}^{h}+{1-\hbox{sign}(x)\over 
       2}\widehat{\sigma}_{II}^{h}\right)+
       +\sum_{l\geq 
	   2}D\widehat{\Phi}^{s,w}_{I,k;I,l}\star\widehat{\rho}_{I,l}^{w}+\ldots\nonumber\\
	   \widehat{\rho}_{II,k}^{s}+\widehat{\rho}_{II,k}^{s,h}&=&e^{-(k-1/2)|x|}
		  \left({1+\hbox{sign}(x)\over 
		  2}\widehat{\sigma}_{II}^{h}+{1-\hbox{sign}(x)\over 2}\widehat{\sigma}_{I}^{h}\right)+
		  +\sum_{l\geq 
		      2}D\widehat{\Phi}^{s,w}_{II,k;I,l}\star\widehat{\rho}_{I,l}^{w}
		      \end{eqnarray}
where each time the dots indicate the obvious sum over the remaining 
terms $(I,s),(II,w),(II,s)$, and the dressed kernels are obtained by 
shifts
\begin{equation}
    D\Phi_{I,n;I,p}^{w,w}=\Phi_{I,n-1;I,p-1}^{s,s}\ldots
    \end{equation}

    The new system bears a lot of resemblance with the old one. In 
    fact, up to a relabelling of the roots, it is identical for what 
    concerns the interactions between the densities $\rho$ of 
    pseudoparticles.
    
 This is fully expected, since the dynamics of the $\rho$ excitations is determined entirely by the symmetries of the model - here, $osp(2|2)$ super Lie algebra. 
 
 Now for the interaction between the sigma densities. In the case of ordinary Lie algebras - say $sl(3)$, of rank two like our present TBA - the holes in $\sigma_{I,II}$ distributions would be associated with particles living in the representations corresponding to the first top or bottom node of the TBA diagram. Oddly enough, while this would be fine if we were considering the second nodes on the diagram (which correspond, as we have seen earlier, to $3,\bar{3}$), there is no $osp(2|2)$ representation corresponding to the first nodes individually! A quick calculation shows that these representations should have dimension $\sqrt{4}=2$ indeed, and this is not possible in $osp(2|2)$. This might indicate that the full $osp(2|2)$ symmetry is not present in the S matrix, but maybe only a sub $sl(2)\times sl(2)$. 
 
 However, it should also be clear that the kernels for $I-I$ and $I-II$ interactions are also quite ill behaved, as their Fourier transforms  both diverge at the origin:
 \begin{eqnarray}
     {1\over i}{d\over d\theta}\ln Z_{I,I}={1\over i}{d\over d\theta}\ln 
     Z_{II,II}=-
     {1\over\pi}\int dx {e^{-|x|}\over 2\sinh |x|} e^{-i\theta 
     x/\pi}\nonumber\\
     {1\over i}{d\over d\theta}\ln Z_{I,II}={1\over i}{d\over d\theta}\ln 
	Z_{II,I}=
	{1\over\pi}\int dx {1\over 2\sinh |x|} e^{-i\theta 
	x/\pi}\label{Zs}
	\end{eqnarray}
(here, $\theta$ is the usual physical rapidity, obtained from the lattice rapidity through a rescaling). We believe this divergence  is related with the non-compactedness of the target space in the UV limit; this will be discussed in more details in the next section. 

Meanwhile, it is intriguing to observe that, if we were to define new excitations by binding a hole in the distribution of real wide roots of type I and a hole in the distribution of wide roots of type II {\sl at the same rapidity}, the Fourier transform of the derivative of the logarithm of 
the dressed scattering kernel of these new excitations with 
themshelves would be
\begin{equation}
    \widehat{\Phi}_{I+II,I+II}=2\left({1\over 2\sinh |x|}-{e^{-|x|}\over 2\sinh 
    |x|}\right)={e^{-|x|/2}\over \cosh |x|/2}
    \end{equation}
Similarly the kernel of such an excitation with a wide string of type 
I or type II is given by $e^{-(k-1)|x|}$ while the kernel of such an 
excitation with a strange string of type I or II is given by 
$e^{(k-1/2)|x|}$. Hence, if we  forced  holes in the I and 
II distributions to be  bound,  the following system of equations deduced from the lattice 
model would now coincide with the TBA of the previous section provided we identify
\begin{equation}
	2\pi\Phi_{\sigma,\sigma}={1\over i}{d\over d\theta}\ln Z= {1\over 
	\pi}\int_{-\infty}^{\infty}{e^{-|x|/2}\over \cosh(x/2)} 
	e^{-i\theta x/\pi}dx
	\end{equation}
Integrating this quantity we find 
\begin{equation}
    Z=(Z_{I,I}Z_{I,II})^2=\left[{\Gamma\left(1+{i\theta\over 
    2\pi}\right)\Gamma\left({1\over 2}-{i\theta\over 2\pi}\right)\over
    \Gamma\left(1-{i\theta\over 2\pi}\right)\Gamma\left({1\over 
    2}+{i\theta\over 2\pi}\right)}\right]^{2}
    \end{equation}
which in fact is the result for the GN model discussed above. In other words, pairs of holes in the I and II distributions at identical rapidities  in the sigma model scatter exactly with the GN S matrix! There is however no obvious reason why these holes should be paired (and the sigma model does not coincide with the GN model anyhow), so the sigma model S matrix appears as some sort of `split' GN S matrix. More work on the bound states and crossing unitarity would be necessary to give sense to this idea.

We note here that in the case $N>0$, the generic sigma model S matrix gives the correct scattering theory for all values of $N$, while the generic GN S matrix works only for large enough $N$; at small enough $N$, the fundamental fermions are unstable, and the correct S matrix involves instead kinks in the spinor representations. It seems that for $N\leq 0$, the roles of the GN and sigma model are switched again: the generic S matrix for GN seems to still work for $N=0$, while for the sigma model at $N=0$, some sort of `spinor' S matrix seems necessary. The spinor representations of $OSp(2|2)$ are however infinite dimensional, and it is not clear whether they would allow  for a meaningful description of the scattering in the present case.

We note finally that the ground state energy of the $OSp(2|2)$ sigma model can be calculated using the general result (\ref{general}) with $N=0$ and one finds again
\begin{equation}
{F\over L}=-{M^2\over 8}\times 0
\end{equation}
This agrees with the ground state energy for the WZW model perturbed by current current interaction \cite{AlyoshaZ}.

\section{Do the massive flows exist?}

A important remark is that, according to the discussion in \cite{Zirnbauer}, the $OSp(2|2)$ WZW model perturbed by a current current interaction in the massive direction in fact does not exist as a field theory. More precisely, it is established in this reference that the functional integral is not well defined, the topological term leading to strong divergences. 
  
The same holds true, in fact, for the sigma model, as was briefly discussed in \cite{JSarboreal}. 
The action for the supersphere  $OSP(2/2)/OSP(1/2)$ sigma model is obtained as follows. Parameterize the $S^{1,2}$ supersphere as
\begin{eqnarray}
    \phi_{1} &=& \cos\phi\left(1-\eta_{1}\eta_{2}\right)\nonumber\\
    \phi_{2} &=& \sin\phi\left(1-\eta_{1}\eta_{2}\right)
\end{eqnarray}
such that $\phi_{1}^{2}+\phi_{2}^{2}+2\eta_{1}\eta_{2}=1$. The action is then
\begin{equation}
    S={1\over g}\int 
    {\rm d}^2x \, \left[\left(\partial_{\mu}\phi\right)^{2}(1-2\eta_{1}\eta_{2})+2\partial_{\mu}\eta_{1}\partial_{\mu}\eta_{2}
    -4\eta_{1}\eta_{2}\partial_{\mu}\eta_{1}\partial_{\mu}\eta_{2}\right]\label{sigmaact}
\end{equation}
where $\phi$ is compactified, $\phi\equiv \phi+2\pi$, and conventions such that the Boltzmann weight is $e^{-S}$.  The 
spontaneously broken symmetry phase, with the theory being 
free in the UV and massless in the IR, corresponds to a {\sl 
positive} coupling constant  $g$, since for the $O(N)$ model, the beta function is 
\begin{equation}
{dg\over dl}=\beta(g)=(N-2)g^2
\end{equation}
and the same holds for $OSp(n|2m)$ sigma models with $N\equiv n-2m$. When $g$ is positive, the theory  flows to weak coupling,
and critical properties are described 
by the $g\rightarrow 0$ limit. A rescaling and 
relabeling brings the action into the form
\begin{equation}
    S=\int 
    {\rm d}^2x \, \left[\left(\partial_{\mu}\phi\right)^{2}
    (1-2g\eta_{1}\eta_{2})+2\partial_{\mu}\eta_{1}\partial_{\mu}\eta_{2}
    -4g\eta_{1}\eta_{2}\partial_{\mu}\eta_{1}\partial_{\mu}\eta_{2}\right]
\end{equation}
with now $\phi\equiv\phi +{2\pi\over\sqrt{g}}$. As $g\rightarrow 0$, we 
thus obtain a symplectic fermion and a non-compact boson.

Note that the presence of  a 
continuous spectrum in this $osp(2/2)$ version of the problem
has a physical origin in the fact 
that the symmetry is spontaneously broken, and thus correlation 
functions of order parameters have no algebraic decay (though  they 
have non trivial  logarithmic behaviour). 

Now the massive region meanwhile corresponds to the theory flowing to strong coupling, and $g$ negative. Except in the $OSp(1|2)$ case, where there is a single bosonic field, and the theory can be reformulated entirely in terms of the symplectic fermions, in the $OSp(2|2)$ case in particular, a negative coupling constant leads to a term $\exp\left[ +\int d^2x (\partial_\mu\phi)^2\right]$ in the Boltzmann weight, and thus to a theory that is not well defined from the path integral point of view. We note that trying to define theories in a region where the path integral is not defined is a task that has been considered in string theory for some time. An example is the so called ``time like Liouville theory'', discussed for instance in \cite{Strominger} and \cite{FredenVolker}. 

So what  happens at $g$ negative and the ``massive'' region is not entirely clear. The model defined by the action (\ref{sigmaact}) is  unstable, a fact that probably manifests itself by a spectrum that becomes unbounded from below, and hence level crossing right at $g=0$. 

The point can be illustrated more concretely in the case of the 
 $OSp(4|2)$ GN model. In the papers \cite{CanduSaleur} and \cite{Volkeretal} it was found that, for the sign of the $O(N)$  GN coupling which corresponds  to the relevant direction in the case $N>2$, the model is gapless and exhibits a line of fixed points with central charge $c=1$.  Meanwhile, for the opposite sign of the coupling, something very different occurs. This can best be seen by contemplating the formula giving the exact values of the conformal weights and thus the scaled gaps on the critical line
 (for the theory with Neumann boundary conditions), as a function of the Young diagrams for the $OSp(4|2)$ representations. Writing the weights at the free point as 
 \begin{eqnarray}
 h_{\hbox{free}}&=&{n_1^2\over 2}+{n_2^2\over 2}+{b-2\over 2},~~b\geq 2\nonumber\\
 h_{\hbox{free}}&=&{n_1^2\over 2},~~b=0,1
 \end{eqnarray}
for a Young diagram with shape $\lambda=n_1n_21^{b-2}$,  the exact formula  \cite{CanduSaleur,Volkeretal} is 
 \begin{equation}
 h=h_{\hbox{free}}+gC
 \end{equation}
 where $C$ is the Casimir of the representation
 \begin{equation}
 C={(n_1+n_2-1)^2\over 2}+{(n_1-n_2+1)^2\over 2}-(b-1)^2
 \end{equation}
 While for $g>0$ the ground state remains in the identity sector, for $g<0$,  the weights of the 
 fully antisymmetric representations $1^p$  (for instance) become arbitrarily large and negative for $p=O(1/\sqrt{g})$ as soon as $g$ is turned on: an infinity of crossings occurs, and the new ground state has nothing to do with the one at $g<0$. 
  
On the other hand, we could also restrict to the  $O(2)$ subsector of the theory. There, the partition function is the one of the free boson with Neumann boundary conditions, and nothing happens - since all the representations which could cross the ground state are eliminated when going from $OSp(4|2)$ to the sub $O(2)$. The properties of the ground state, analytically continued in this phase, are well behaved, but describe a subset of the theory which is very high in energy. The equivalent observation in the $OSp(2|2)$ GN model would correspond to the sector with periodic boundary conditions, where the partition function remains equal to unity and $c=0$; this sector might be very high in energy with respect to the true ground state in the would be massive phase.

 The $OSp(2|2)$ sigma model can be studied somewhat in the same spirit by using perturbation theory. Near the weak coupling fixed point, the massive direction formally corresponds to negative $g$, and, if we for a minute do not worry about the meaning of the path integral, we can calculate the evolution of the hamiltonian eigenvalues perturbatively. Symmetric representations are particularly easy to handle, since they can also be obtained through the minisuperspace analysis. The eigenvalues are simply proportional to those of  the Laplacian on the supersphere:
\begin{equation}
H=g\Delta_{S^{-1}}
\end{equation}
and of the form $E=g n(n-2)$, $n$ an integer. They all collapse at $g=0$, and there is an infinity of level crossings at that point when going from $g$ positive to $g$ negative. 

We note that, if this scenario is correct, the point $g=0$ must be considered as a `first order critical point''. This is a point where the theory is conformal invariant, but which is at the same time a point where level crossing occurs, such that the analytic continuation of the ground state  becomes a very highly excited state in the low temperature phase. In general, first order phase transitions are not associated with conformal field theories, but several examples of first order critical points are known. These involve crossings in finite size, and are only possible because of the non unitarity. 
Maybe the best studied example corresponds to the thermal perturbation of the antiferromagnetic Potts model on the square lattice, as discussed in \cite{JS}.

Meanwhile, we have a lattice regularization (the staggered spin chain) and a TBA that seem perfectly stable. In particular, we have not found, by exploring solutions which are dissymmetric in the two types of  roots I and II, any indication that the ground state of the chain is not given by the foregoing results. It looks thus as if our results  describe  some sort of continuation of the unstable theory, which now behaves as an (almost) ordinary massive theory. It would of course be interesting to explore this issue in more details.

 \section{The q-deformed case}

 We have so far discussed only the case of twisted boundary conditions for the fermions, corresponding to the ``effective'' central charge of our non unitary theories. To get the true central charge, one must consider a twisted TBA, by inserting (complex) chemical potentials for the kinks in the scattering. For the $OSp(2|2)$ GN model, we do not expect that much interesting will remain. The central charge will be zero in the UV,  it is zero in the IR for the massive flow, and probably remains zero all the way. 
 
 The situation is more interesting for the supersphere sigma model, since now $c=-1$ in the UV. To get the calculation of the true central charge under control, it is convenient to consider the q deformed version of our problem. 
 
 In terms of lattice model, the q-deformation is obtained by taking the four dimensional solution of the YB equation based on $U_qOSp(2|2)$ \cite{Maassarani}. 
 
 Setting $q=e^{i\xi}$ we have
 the Bethe equations for a chain with $L$ sites
\begin{eqnarray}
\left({\sinh{1\over 2}(U_j-i\xi )\over \sinh{1\over 2}(U_j+i\xi)}\right)^L=\prod_\beta {\sinh{1\over 2}(U_j-\Gamma_\beta-2i\xi)\over\sinh{1\over 2}
(U_j-\Gamma_\beta+2i\xi)}\nonumber\\
\left({\sinh{1\over 2}(\Gamma_\beta-i\xi)\over \sinh{1\over 2}(\Gamma_\beta+i\xi )}\right)^L=\prod {\sinh{1\over 2}(\Gamma_\beta-U_j-2i\xi)\over\sinh{1\over 2}
(\Gamma_\beta-U_j+2i\xi)}
\end{eqnarray}
(the isotropic limit is recovered by setting $U_j=2\xi u_j, \Gamma_\beta=2\xi \gamma_\beta$ and $\xi\to 0$. 
The energy of the chain is meanwhile
\begin{equation}
E\propto -\left(\sum_k  {\sin\xi\over \cosh U_j-\cos\xi}+\sum_\beta
 {\sin\xi\over \cosh \Gamma_\beta-\cos\xi}\right)
 \end{equation}
In the antiferromagnetic regime $\xi\in [0,{\pi\over 2}]$, the ground state is obtained by filling up the seas $y_I,y_{II}$ with real roots. With anti-periodic boundary conditions for the fermionic degrees of freedom (the NS sector), the central charge reads $c_{\hbox{eff}}=2$. 
 
The most natural is to observe that the XXZ sub component of the spectrum (that is, the component given by $y_I=y_{II}$) can be twisted to obtain a $c<1$ theory described by a Coulomb gas with parameter $\alpha_0$. To do so we impose boundary conditions twisted by $e^{2i\pi \phi}$ on the spins, $\phi={\xi\over\pi}$
resulting in the central charge for the XXZ chain
\begin{equation}
c=1-{6(\xi/\pi)^2\over 1-(\xi/\pi)}
\end{equation}
We now parametrize 
\begin{equation}
{\xi\over\pi}={1\over 2l+3}
\end{equation}
(usually one would set $\xi/\pi=1/(m+1)$) so the central charge of the $OSp$ chain becomes
\begin{equation}
c_{\hbox{twist1}}=2\left[1-{6\over (2l+2)(2l+3)}\right]
\end{equation}
Meanwhile, we recall that the effective central charges (ie, describing the ground state scaling in the NS sector) of the $OSp(2|2)$ and $OSp(1|2)$ WZW theories at level $l$ read
\begin{eqnarray}
OSp(2|2)_l;~~c=0;~~c_{\hbox{eff}}={6l\over l+1}\nonumber\\
OSp(1|2)_l;~~c={2l\over 2l+3};~~c_{\hbox{eff}}={8l\over 2l+3}
\end{eqnarray}
It is easy to check that 
\begin{equation}
c_{\hbox{twist1}}={6l\over l+1}-{8l\over 2l+3}
\end{equation}

This suggests that the continuum limit of the spin chain is described,  for $l$ integer, by coset models  $OSp(2|2)_l/OSp(1|2)_l$ and suitable interpolations in between.

When one goes from the XXZ to the minimal models, a set of electric charges have to be introduced, which are multiple of the fundamental charge $e_0={\gamma\over\pi}$. The particular choice $e=(l+1)e_0$ gives the conformal weight, with respect to the central charge of the minimal model, 
$$
h_t={(l+1)^2-1\over 4(2l+2)(2l+3)}
$$
This leads to an effective central charge for the $OSp$ theory
\begin{equation}
c_{\hbox{twist2}}=c_{\hbox{twist1}}-24\times 2\times h=-{2l\over 2l+3}
\end{equation}
which is the true central charge of the coset model. The phase shift to obtain this central charge is thus
$\exp[2i\pi (l+1)/(2l+3)]$; it becomes a pure $e^{i\pi}$ in the $l\rightarrow\infty$ limit, where the central charge $c_{\hbox{twist2}}$ becomes equal to $-1$. 

A way to obtain the flow of the central charge in the supersphere sigma model is thus to take the flow of the running dimension in the minimal models $M_{2l+2,2l+3}$ in the limit $l\rightarrow \infty$. 

Now this is particularly simple. Indeed, we observe that $h_t=h_{l+1,l+1}$, and the running dimension of operators $\Phi_{rr}$ in minimal models is the easiest to obtain, since it follows from a simple twist on the
basic sine-Gordon equations. This can be done in the context of the TBA, as discussed in \cite{PaulandI}. In the notations of this paper, one must set $\gamma={1\over 2}$, so the soliton fugacities are $\exp[\pm i\pi(2l+3)/2]$. One can also use the DDV formalism, and for instance the formulas in \cite{monstron}. The basic equations reads
\begin{eqnarray}
f(\theta)=imR\sinh\theta+i\alpha+\int_{C_1}\Phi(\theta-\theta')\ln\left(1+e^{f(\theta')}\right)d\theta'
+\int_{C_2}\Phi(\theta-\theta')\ln\left(1+e^{-f(\theta')}\right)d\theta'
\end{eqnarray}
where $\Phi$ is the SG kernel for the corresponding value of the sine-Gordon coupling, ${\beta_{\rm SG}^2\over 8\pi}=1-{1\over 2l+3}={2l+2\over 2l+3}$, $C_1$ and $C_2$ are contours slightly below and above the real axis, and the effective central charge follows from
\begin{eqnarray}
c(mR,\alpha)={3imR\over \pi^2}
\left[\int_{C_1}\sinh\theta\ln\left(1+e^{f(\theta)}\right)d\theta+\int_{C_2}\sinh\theta\ln\left(1+e^{-f(\theta)}\right)\right]d\theta
\end{eqnarray}
In the present case we need $\alpha={\pi\over 2}$ exactly. The flow now should go from $c=-{l\over 2l+3}$ in the UV to $c=0$ in the IR. In the limit $l\rightarrow\infty$, this becomes a flow from $c=-1/2$ up to $c=0$. Multiplying the result by a factor of two gives the desired result for the $OSp(2|2)$ case.

We now get back to the untwisted sector with $c_{\hbox{eff}}=2$, and discuss the finite size spectrum. Excitations are obtained by changing the number of roots or shifting the roots. Parametrize the numbers of roots as
\begin{equation}
r_{I,II}={L\over 2}-n_{I,II}
\end{equation}
and suppose the Bethe integers are submitted to a global shift given by $m_{I,II}$. One then finds (similar formulas appear in \cite{GalleasMartins})
\begin{eqnarray}
x_{(n_I,n_{II}),)(m_I,m_{II})}&=&\left(1-{\xi\over\pi}\right) \left({n_I+n_{II}\over 2}\right)^2+{1\over 1-{\xi\over \pi}}
 \left({m_I+m_{II}\over 2}\right)^2\nonumber\\
&+& {\xi\over\pi} \left({n_I-n_{II}\over 2}\right)^2+{1\over{\xi\over \pi}}
 \left({m_I-m_{II}\over 2}\right)^2\label{FSS}
 \end{eqnarray}

 For reference it is useful to recall the finite size spectrum of the XXZ antiferromagnetic chain with $\Delta=-\cos\xi_0$ :
 \begin{equation}
 x_{n,m}=\left(1-{\xi_0\over\pi}\right){n^2\over 2}+{1\over 1-{\xi_0\over\pi}}{m^2\over 2}
 \end{equation}
 We observe first that excitations with $n_I=n_{II}=n$ and $m_I=m_{II}=m$ in the $OSp$ chain are equal to twice the excitations in the XXZ chain, a result that can be checked directly at the level of the Bethe equations. 
 
 The $OSp$ quantum numbers are given by 
 \begin{eqnarray}
 b={n_I-n_{II}\over 2}\nonumber\\
 S^z={n_I+n_{II}\over 2}
 \end{eqnarray}
 so the sub-component of the spectrum arising from excitations $m_I=m_{II}=0$ reads
 \begin{equation}
 x_{(n_I,n_{II})(0,0)}=\left(1-{\xi\over\pi}\right) (S^z)^2+ {\xi\over\pi} b^2
  \end{equation}
We see on (\ref{FSS}) that  the finite size spectrum is determined by two bosonic fields with related coupling constants.  When $\xi\to 0$, one of the coupling constants vanishes, and the theory reduces to a non compact boson and a compact one at the Dirac radius.  

We now consider, like in the case $\xi=0$, a staggering of the $OSp(2|2)_q$ chain (adding $\pm i\Lambda$ to $U,\Gamma$ in the Bethe equations), in order to induce a massive deformation.  We now use Fourier transforms defined as 
\begin{equation}
\hat{f}(k)=\int {dU\over 2\pi} e^{iUk/2\xi} f(U)
\end{equation}

The equivalent of equations (\ref{systemi}),  is now 
\begin{eqnarray}
\sigma_I+\sigma_I^h={\cos k\Lambda/2\xi \over 2\cosh k/2}+\Phi_{I,I}\sigma_I^h+\Phi_{I,II}\sigma_{II}^h+\ldots\nonumber\\
\sigma_{II}+\sigma_{II}^h={\cos k\Lambda/2\xi \over 2\cosh k/2}+\Phi_{II,I}\sigma_I^h+\Phi_{II,II}\sigma_{II}^h+\ldots
\end{eqnarray} 
where we have not written the interaction with the other strings, and the kernels are given by 
\begin{eqnarray}
\Phi_{I,I}=\Phi_{II,II}=-{\cosh k({\pi\over \xi}-2)-1\over 2\sinh k({\pi\over \xi}-1)\sinh k}\nonumber\\
\Phi_{I,II}=\Phi_{II,I}={\cosh k({\pi\over \xi}-1)-\cosh k\over 
2\sinh k({\pi\over \xi}-1)\sinh k}
\end{eqnarray}

The dominant pole in the source term is a $k=i\pi$, so the mass term goes as $M\propto e^{-\Lambda\pi/2\xi}$, and the free energy will be a function of the product $M/T$ as usual. We have then to match analytical properties of the free energy deduced from the lattice equations with the relation between the physical mass and the bare mass (related with the staggering). 
A few trials seem to leave only one option: a perturbation of the type
\begin{equation}
A=\int d^2x {1\over 2}\sum_{i=1,2}\partial_\mu\phi_i\partial^\mu\phi_i+\lambda \cos\beta_1\phi_1e^{i\beta_2\phi_2}+\mu e^{-2i\beta_2\phi_2}\label{conjecture}
\end{equation}
with the constraint 
\begin{equation}
\beta_1^2+\beta_2^2=4\pi\label{constr}
\end{equation}
Eq. (\ref{constr}) follows from the finite size lattice spectrum, where  pure $\phi_i$ excitations are obtained with $m_{I}=m_{II}=0$ and $n_I=\pm n_{II}$. To proceed, we observe that the action in (\ref{conjecture}) leads to a free energy expanding in even powers of $\lambda^2\mu L^{4-\beta^2_2/\pi}=\lambda^2\mu L^{\beta_1^2/\pi}$ (where we used (\ref{constr})). It follows that 
\begin{equation}
M\propto (\lambda^2\mu)^{\pi\over \beta_1^2}
\end{equation}
Meanwhile,  the perturbation induced by the staggering on the microscopic hamiltonian is proportional to $e^{-\Lambda}$, so we get the basic identification ${2\pi\over \beta_1^2}={\pi\over 2\xi}$ or 
\begin{eqnarray}
{\beta_1^2\over 8\pi}&=&{\xi\over 2\pi}\nonumber\\
{\beta_2^2\over 8\pi}&=&{1\over 2}-{\xi\over 2\pi}\label{param}
\end{eqnarray}
This is in the finite size spectrum indeed. Pure $\phi_1$ excitations 
correspond to $n_I=-n_{II}$, pure   $\phi_2$ excitations to $n_I=n_{II}$. When $\xi\to 0$, we get in the action, to leading order, $e^{i\sqrt{4\pi}\phi_2}$ and $e^{-2i\sqrt{4\pi}\phi_2}$, that is, an $a_2^{(2)}$ action, which is known \cite{Birgit1} to  be equivalent to the pure fermion terms in (\ref{sigmaact}) after twisting (to et $c_{\hbox{eff}}=2$). The $\phi_1$ field, meanwhile, must correspond to the non compact $\phi$ field in (\ref{sigmaact}). We recover the expectation that the non compact direction corresponds to excitations $n_I\neq n_{II}$ then. 

We note now that the model (\ref{conjecture}) is a complex version of the model dubbed ${\cal C}_1^{(1)}$ in \cite{Fateev1}, and well known to be integrable. Our analysis gives as a by-product the free energy of this theory: similar arguments as in earlier sections about the sector symmetric under $I\leftrightarrow II$ exchange shows that 
\begin{equation}
{F\over L}({\cal C}_1^{(1)}-\hbox{ complex})=2{F\over L}\left(SG,{\beta_{SG}^2\over 8\pi}={\beta_2^2\over 4\pi}\right)
\end{equation}
where (\ref{param}) must be used to obtain $\beta_1$. We do not know how such a relation could be proven directly. 

Another intriguing observation is that the kernels for the $I-I$ and $I-II$ scattering, which read (compare with the isotropic case (\ref{Zs}) recovered by letting $\xi\to\infty$ and identifying $k\equiv x$)
\begin{eqnarray}
{1\over i}{d\over d\theta}\ln Z_{I,I}=-{1\over \pi}\int dk {\cosh ({\pi\over \xi}-2)k-1\over
2\sinh({\pi\over \xi}-1)k\sinh k }e^{-i\theta k/\pi}\nonumber\\
{1\over i}{d\over d\theta}\ln Z_{I,II}={1\over \pi}\int dk {\cosh ({\pi\over \xi}-1)k-\cosh k\over
2\sinh({\pi\over \xi}-1)k\sinh k }e^{-i\theta k/\pi}\label{ZCSG}
\end{eqnarray}
(here $\theta$ is the physical rapidity, $\theta={\pi\over 2\xi} U,{\pi\over 2\xi}\Gamma$ obey the relation
\begin{equation}
{1\over i}{d\over d\theta}\ln Z_{I,I}Z_{I,II}=-{1\over \pi}\int dk{\sinh ({\pi\over 2\xi}-1)k\over 2\sinh ({\pi\over 2\xi}-{1\over 2})k\cosh {k\over 2}}e^{-i\theta k/\pi}\label{ZSG}
\end{equation}
This is exactly the sine-Gordon kernel at 
\begin{equation}
{\beta_{SG}^2\over 8\pi}=1-{\xi\over \pi}
\end{equation}
A more detailed analysis shows that combining holes of type $I$ and type $II$ gives excitations whose scattering is exactly given by the S matrix for the q-deformed GN theory in \cite{BassiLeclair}. It is also possible to use the limit $q\to 1$ to try to make sense of the diverging kernels encountered in the isotropic case. This will be discussed elsewhere. 

Note finally that when $\gamma={\pi\over 2}$, the theory seems entirely free: no interactions remain in the Bethe ansatz, and $Z_{I,I}=Z_{I,II}=1$. This corresponds meanwhile to the case $\beta_1=\beta_2=\sqrt{2\pi}$, which will also be discussed elsewhere. 

 %
 %
 
 \section{Application: flow in the Random Bond Ising model}

In a very interesting paper, Mussardo et al. \cite{Giuseppe} argued  that the flow 
into the random bond Ising model could be described by an $O(N)$ 
massless scattering in the `limit' $N\rightarrow 0$. The $S$ matrix proposed in \cite{Giuseppe}
involves  left movers in a  copy of the fundamental representation 
of $O(N)$, right movers in another copy, with 
$S_{RR}=S_{LL}=S_{RL}$. This is the obvious structure one might guess 
from experience in the flow from $O(3)$ to $SU(2)_{1}$ WZW model. 
Such an S matrix structure has also  been considered by P. Fendley in \cite{Fendley}. There, the question being studied was the existence of a massless flow into the $O(2N)_1$ WZW model. The S matrix proposed had a similar structure, $S_{RR}=S_{LL}=S_{GN}$ (the case of general $N$ is considered, with particles in the fundamental representation). The S matrix for $S_{LR}$ in \cite{Fendley} required some additional CDD factors to get rid of spurious poles in the LR channel. It was then  found to  describe a flow from the $O(2N)/O(N)\times O(N)$ sigma model at $\theta=\pi$ into the $O(2N)_1$ WZW model. 

Of course the regime studied in \cite{Fendley} concerns $N>2$, and has a different behavior from the one in \cite{Giuseppe} since relevant and irrelevant directions are switched as $N$ crosses two. We thus cannot use the results  in \cite{Fendley} analytically continued in $N$ to see what happens in the random bond Ising model. 

The natural strategy to make progress is then to  turn to a supersymmetric version \cite{Bernard} of the argument in \cite{Giuseppe}, replacing the $N\to 0$ limit by $OSp(2|2)$. The goal is then to find a RG trajectory ending up at the $OSp(2|2)_1$ WZW model in the IR, and originating form a non trivial UV fixed point. The proposal in \cite{Giuseppe} immediately extends to $S_{RR}=S_{LL}=S_{RL}$ given by the S matrix in section 4, with $R$ particles having dispersion relation $e=p={M\over 2}~e^{\theta_R}$, and $L$ particles $e=-p={M\over 2}~e^{-\theta_L}$. 

Note that, in contrast with the case discussed in \cite{Fendley}, no CDD factor is a priori necessary, since there are no poles in the GN S matrix any longer.

\begin{figure}
 \centering
     \includegraphics[width=2in]{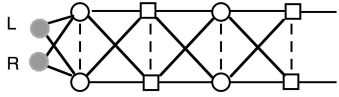}
 \caption{TBA for the susy version of the massless flow into the 
 random bond Ising model. }
 \label{tbadiag4}
 \end{figure}

\begin{figure}
  \centering
      \includegraphics[width=2in]{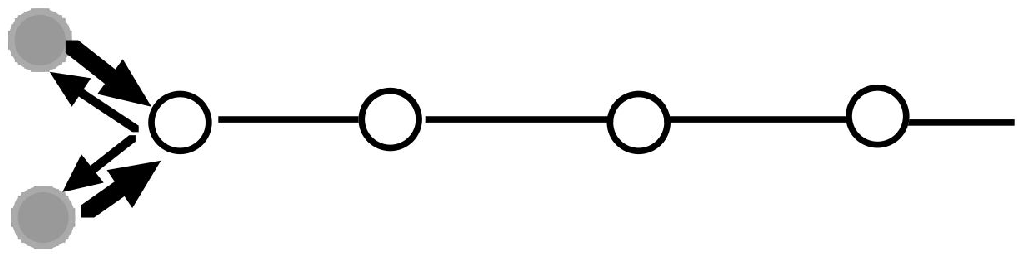}
  \caption{TBA for the susy version of the massless flow into the 
  random bond Ising model: an equivalent form after folding. }
  \label{tbadiag5}
  \end{figure}

The TBA in the Neveu Schwartz sector is thus obtained by an immediate generalization of the results in 
the previous section, where instead of having a massive node on the 
left, we have a L and a R moving node, identically coupled to the 
body of the diagram, with no coupling between them.   See  figure \ref{tbadiag4},\ref{tbadiag5} and the appendices. 
It is clear that this TBA converges to the correct theory in the IR. 
The question is, where does it originate from?

Let us write the TBA explicitely. We give to the two grey nodes 
labels $0,\bar{0}$, and to the other nodes labels $1,2,\ldots$. We 
have thus
\begin{eqnarray}
    \epsilon_{0}(\theta)&=&{M\over 2}e^{\theta}-2 Ts\star 
    \ln\left(1+e^{-\epsilon_{1}/T}\right)\nonumber\\
     \epsilon_{\bar{0}}(\theta)&=&{M\over 2}e^{-\theta}-2T s\star 
	\ln\left(1+e^{-\epsilon_{1}/T}\right)\nonumber\\
	\epsilon_{1}&=& - Ts\star 
	\left[\ln\left(1+e^{-\epsilon_{0}/T}\right) +
	\ln\left(1+e^{-\epsilon_{\bar{0}}/T}\right)
	+	       \ln\left(1+e^{-\epsilon_{2}/T}\right)\right]\nonumber\\
	       \epsilon_{2}&=& -Ts\star 
		      \left[ \ln\left(1+e^{-\epsilon_{1}/T}\right)+
		       \ln\left(1+e^{-\epsilon_{3}/T}\right)+\ldots\right]
		       \end{eqnarray}
		   and
\begin{equation}
    {F\over L}=-{1\over 2\pi}\int {M\over 2}e^{\theta}\ln\left(1+e^{-\epsilon_{0}/T}\right)
    +{M\over 2} e^{-\theta}\ln\left(1+e^{-\epsilon_{\bar{0}}/T}\right)
	\end{equation}
	where $s={1\over 2\pi\cosh\theta}$ and $\star$ denotes convolution.

This is quite similar to the TBA for the $O(3)$ sigma model at $\Theta=\pi$; in fact, the only difference is some factors of two in the coupling to the source nodes.

The game now consists in solving these equations numerically (after some truncation) and studying the IR limit. This is a long technical story delineated in the appendix. The main conclusion is that the TBA develops a singularity at some value of $M/T$, and that this value is very large, and probably infinite. Such intriguing behavior was studied in  \cite{MussardoSimon} for instance, where it was interpreted as a sign that the theory one is trying to define encounters some sort of phase transition. For the examples discussed in \cite{MussardoSimon}, the singularity remained at a finite value of $M/T$, so the vicinity of the IR fixed point at least was well defined. In our case, it seems the singularity might occur right near the IR fixed point, so the WZW theory sits on the border of a domain of instability {\sl in the massless region}. Meanwhile, we have found that adding a CDD factor in the LR scattering seems to give a meaningful TBA, provided one chooses 
\begin{equation}
f=\tanh^2(\theta/ 2+i\pi/4)
\end{equation}
Note that this factor has a double pole at $\theta=i/2$, which leads to a `monstron' with complex mass 
\begin{equation}
M_m=Me^{i\pi/4}
\end{equation}
similar to the one encountered in the study of the massless flow from dilute to dense polymers \cite{FSZ,monstron}. The TBA now gives $c_{\hbox{eff}}=1$ in the UV, but it is not clear whether one should add to it a monstron contribution \cite{Dorey},\cite{monstron}. We hope to present a more detailed study of this question elsewhere.

\section{Conclusion}

On the positive side, we have, in this paper: 

\begin{itemize}
\item{} built the basic TBA for $OSp(2|2)$ symmetric integrable models
\item{} found evidence that the S matrix of \cite{BassiLeclair} for the GN model is the correct one
\item{} uncovered an intriguing duality between different quantum field theories, somewhat generalizing to the massive case the results from \cite{CanduSaleur,Volkeretal}
\item{} uncovered yet another duality between the (q-deformation of) the $OSp(2|2)$ sigma model and some integrable two bosons theories
\item{} shown that the proposal of \cite{Giuseppe}  for the flow into the random bond Ising model is not correct 
\end{itemize}

\noindent We have also  uncovered some troubling features that seem very generic. In particular

\begin{itemize}
\item{} the S matrix of the $OSp(2|2)$ sigma model appear to be a very  intriguing object, where some of the usual principles might not apply. It is not known up to now, even though the TBA has been extracted from a lattice regularization. It is also not known what kind of `sausage deformation' of the sigma model corresponds to the q-deformation of the lattice equations.
\end{itemize}

It also remains to see how our analysis might fit into the more general scheme recently developed in the AdS/CFT enterprise. We note in this respect that the problems discussed in this paper could a priori be tackled as well with another form of the Bethe ansatz, corresponding to choosing a different Dynkin diagram. This question was briefly alluded to in \cite{JSarboreal}. We leave this question to further study, and simply notice that the Y system for the GN model reads (setting as usual $Y=e^{-\epsilon/T}$)
\begin{eqnarray}
Y_0(\theta-i\pi/2)Y_0(\theta+i\pi/2)&=&(1+Y_1(\theta))^2\nonumber\\
Y_{\bar{0}}(\theta-i\pi/2)Y_{\bar{0}}(\theta+i\pi/2)&=&(1+Y_1(\theta))^2\nonumber\\
Y_1(\theta-i\pi/2)Y_1(\theta+i\pi/2)&=&(1+Y_0(\theta))(1+Y_{\bar{0}})(\theta))(1+Y_2(\theta))\nonumber\\
Y_i(\theta-i\pi/2)Y_i(\theta+i\pi/2)&=&(1+Y_{i-1}(\theta))(1+Y_{i+1})(\theta))
\end{eqnarray}
This differs from the one for the $O(3)$ model only by a power of two appearing in the first two equations. It would be most interesting to match this Y system with some of the general results uncovered recently \cite{Tsuboigene, Kazakov,Vieira}.

We would like to end this paper with a speculation. We have identified the q-deformed $OSp(2|2)$ sigma model with a complex version of the ${\cal C}_1^{(1)}$ theory discussed in \cite{Fateev1}. Now, for the real case, V. Fateev conjectures that the ${\cal C}_1^{(1)}$ theory is dual to the complex sinh-Gordon model (CSG). Extending (maybe too naively) his conjecture to the complex case, we find that our q-deformed $OSp(2|2)$ sigma model should be equivalent  to the complex sine-Gordon model with action
\begin{equation}
A={1\over 2} \int {\partial_\mu \chi\partial_\mu\bar{\chi}\over 1-{\pi^2\over \pi-\gamma}|\chi|^2}-m_0^2|\chi|^2+\hbox{ counterterms}
\end{equation}
In particular, we obtain that the pure $OSp(2|2)$ case should correspond to 
\begin{equation}
A={1\over 2} \int {\partial_\mu \chi\partial_\mu\bar{\chi}\over 1-\pi|\chi|^2}-m_0^2|\chi|^2+\hbox{ counterterms}
\end{equation}
Of course, the coupling is now so big in the CSG that it is not clear what the action means. Yet,  another important observation in favor of this proposal concerns conserved quantities. It is well known that the CSG model admits  independent conserved quantities of arbitrary spin \cite{Bakas}.  On the other hand,  the S matrix encodes conserved quantities \cite{Niedermaier}; for instance, expanding the sine-Gordon S matrix  (\ref{ZSG}), the poles at $k=(2n+1)i\pi$ correspond to the local conserved quantities, which act as  $e^{(2n+1)\theta}$ 
on particles, and thus  are present only for odd spin. Meanwhile, in the conjectured S matrix (\ref{ZCSG}), there are poles at all $k=n\pi$, leading also to conserved quantities acting as $e^{2n\theta}$. These poles disappear in the product $Z_{I,I}Z_{I,II}$ but are otherwise present, as would be required for the CSG. 

 Interestingly, the question of S matrices for the complex sine-Gordon model is far from being settled \cite{CSG,Miramontes}. Our results provide an intriguing light on this question, indicating a spectrum quite different from the semi classical proposals. We will discuss this issue in more details elsewhere.

\bigskip
\noindent{\bf Acknowledgments}: we thank A. Babichenko, D. Bernard,
C. Candu, V. Fateev A. Leclair, S. Lukyanov, G. Mussardo, N. Dorey, N. Read and  V. Schomerus  for
discussions. We also thank F. Essler, H. Frahm, I. Ikhlef and J. Jacobsen  for earlier
collaboration on related matters. One of the authors (B.P.) acknowledges support by the
ESF INSTANS Exchange Grant (ref. number: 1799) 
and also by the Hungarian OTKA grants K75172 and K60040. The other author acknowledges support from the Agence Nationale pour la Recherche.

\section*{Appendix A: Numerical study of the massless TBA}
\renewcommand\thesection{A}
\renewcommand{\theequation}{A\arabic{equation}}
\setcounter{equation}{0}

\subsection{$O(3)$ sigma-model with $\theta=\pi$: a reminder}

\begin{figure}[ht!]
  \centering
  \includegraphics[scale=0.5]{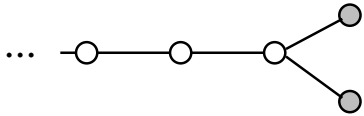}
  \caption{TBA for the $O(3)$ sigma-model with $\theta=\pi$} 
  \label{fig:sigma_pi}
\end{figure}

We first consider this case as a warmup. The infinite set of TBA
equations is well known (see also Fig. \ref{fig:sigma_pi}):
\begin{equation}
\begin{split}
  \eps_0&=r e^\theta/2-s \star \ln\big(1+e^{-\eps_1}\big)\\
 \eps_{\bar{0}}&=r e^{-\theta}/2-s \star \ln\big(1+e^{-\eps_1}\big)\\
 \eps_1&=-s \star \ln\big(1+e^{-\eps_0}\big)-
s \star \ln\big(1+e^{-\eps_{\bar{0}}}\big) 
-s \star\ln\big(1+e^{-\eps_2}\big)\\
 \eps_i&=-s \star \ln\big(1+e^{-\eps_{i-1}}\big)-
s \star \ln\big(1+e^{-\eps_{i+1}}\big) \quad \text{for} \quad
i=2,3,\dots
\end{split}
\label{TBA_sig_pi}
\end{equation}
with $s=\frac{1}{2\pi \cosh{\theta}}$, we have rescaled the
$\epsilon$'s with respect to  the main text, and set $r=M/T$.

The dimensionless free energy per unit length $F(r)$ is
calculated as
\begin{equation*}
  F=-\frac{1}{2\pi}\int  \frac{e^\beta}{2}  \ln\big(1+e^{-\eps_0}\big)
+ \frac{e^{-\beta}}{2}  \ln\big(1+e^{-\eps_{\bar{0}}}\big)
\end{equation*}
In the UV and IR limit it is related to the  central charge as $F=-c/12$.

The above set of equations has to be truncated to be accesible to numerical
investigations. Let us denote the total number of nodes with $N$. The expected  values of the central
charge are 
\begin{equation*}
  c_{UV}=2-\frac{6}{N+2} \quad\quad\quad\quad c_{IR}=1-\frac{6}{(N+2)(N+3)}
\end{equation*}

\begin{figure}
\psfrag{cinIR}{$c_{IR}$}
\psfrag{cinUV}{$c_{UV}$}
\psfrag{log10r}{$\log r$}
\psfrag{F}{$F$}
  \centering
\subfigure[$N=6$]{\includegraphics{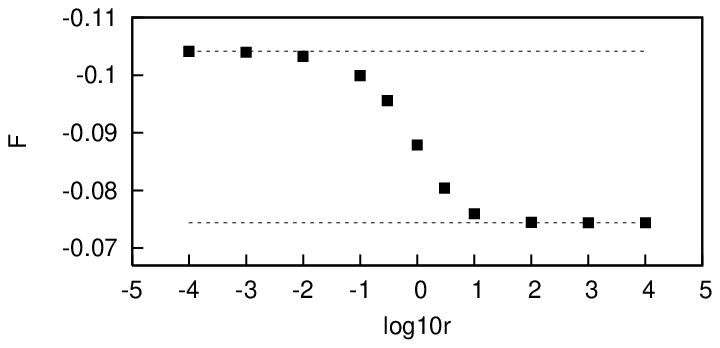}}
\subfigure[$N=10$]{\includegraphics{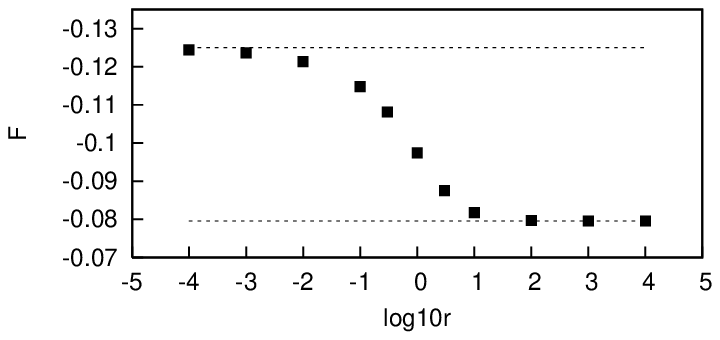}}
\caption{$F(r)$ in the truncated versions of the $\sigma$-model TBA}
\label{sN6}
\end{figure}

The function $F(r)$ is plotted for two different values of $N$ in 
Fig. \ref{sN6}.  

\begin{figure}
  \centering
\psfrag{r0.1}{\small $r=0.1$}
\psfrag{rii1}{\small $r=1$}
\psfrag{ri10}{\small $r=10$}
   \includegraphics{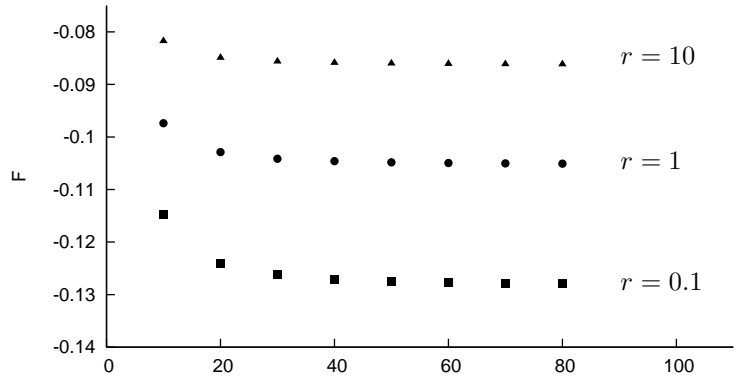}
\caption{$F$ as a function of $N$ for $r=0.1,1,10$ in the $O(3)$ sigma-model
  with $\theta=\pi$}
\label{FofNsigma}
\end{figure}

Convergence as a function of $N$ is clearly rather quick. To see this better, 
consider $F$ as a function of $N$ for different
fixed values of $r$. The results for $r=0.1,1,10$ are shown in
Fig. \ref{FofNsigma}. One can see that there is a well-defined
$N\to\infty$ limit, the leading correction is $O(1/N)$.

\subsection{The conjectured massless flow in the 
  $OSP(2|2)$ case}

The infinite set of TBA equations given in the text becomes:
\begin{equation}
\begin{split}
  \eps_0&=r e^\theta/2-2s \star \ln\big(1+e^{-\eps_1}\big)\\
 \eps_{\bar{0}}&=r e^{-\theta}/2-2s \star \ln\big(1+e^{-\eps_1}\big)\\
 \eps_1&=-s \star \ln\big(1+e^{-\eps_0}\big)-
s \star \ln\big(1+e^{-\eps_{\bar{0}}}\big) 
-s \star\ln\big(1+e^{-\eps_2}\big)\\
 \eps_i&=-s \star \ln\big(1+e^{-\eps_{i-1}}\big)-
s \star \ln\big(1+e^{-\eps_{i+1}}\big) \quad \text{for} \quad
i=2,3,\dots
\end{split}
\label{eq:massless_osp1}
\end{equation}
with $s=\frac{1}{2\pi \cosh{\theta}}$. The only difference compared to
the case of the $\sigma$-model is a factor of 2 in the equations for $\eps_0$ and
$\eps_{\bar{0}}$. However, this factor dramatically changes the
behaviour of the TBA. Similar to the $\sigma$-model, we first study the
truncated systems with $N$ nodes.

 \begin{figure}
  \centering
      \includegraphics[scale=0.5]{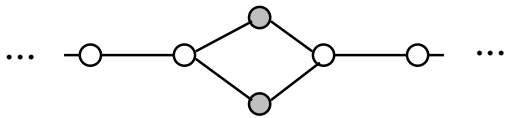}
  \caption{TBA for the susy version of the massless flow into the 
  random bond Ising model: an equivalent form after folding. }
  \label{fig:no_cdd}
  \end{figure}

The unfolded version of the TBA is represented by the graph in
fig. \ref{fig:no_cdd}, which only contains simple links. 
Given this structure of the TBA, only those diagrams can
describe a meaningful flow for which the largest eigenvalue
$\lambda_{\text max}$ of the
incidence matrix is less than 2   \cite{dynkinTBA}. If this condition is satisfied, in
the UV limit one
observes the usual pattern of plateus in the functions $L_i=\log
(1+e^{-\eps_i})$ in the central region  $|\theta|<\log r$. On the other
hand, if $\lambda_{\text max}>2$, then one expects divergences in the
UV limit \footnote{$\lambda_{\text max}$ may be equal to 2, but then this
should hold for every finite $N$. In this case the pseudoenergies may
be infinite in the UV limit, whereas the effective central charge
stays finite.}.  This is the case for the diagram  \ref{fig:no_cdd}, even for
finite $N$. Nevertheless the 
calculation for the central charge is well-defined in the IR and leads to
\begin{equation*}
  c_{IR}=3-3\frac{3N-1}{N(N+1)}
\end{equation*}

\begin{figure}
\psfrag{cinIR}{$c_{IR}$}
  \centering
\subfigure[$N=6$]{\includegraphics{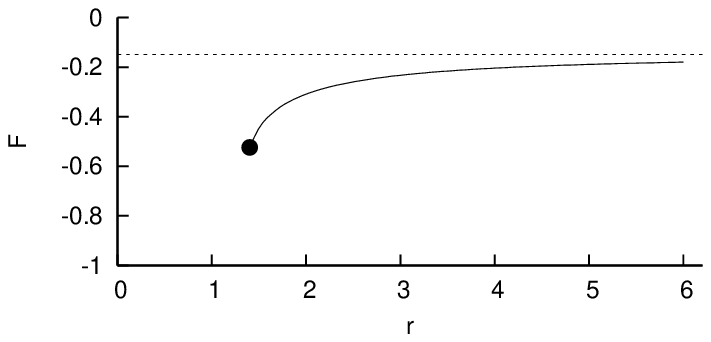}}
\subfigure[$N=10$]{\includegraphics{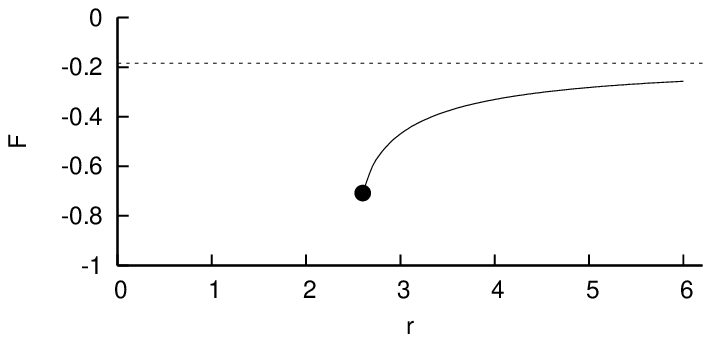}}
\caption{$F(r)$ in the truncated version of the massless TBA
  \eqref{eq:massless_osp1}. The dashed line corresponds to the
  analytical prediction in the IR. The filled circles show the
  critical values of $r$ where the TBA equations become unstable.}
\label{Fr-N6}
\end{figure}

We performed numerical calculations for different values of $N$, the
results for $N=6$ and $N=10$ are shown in Fig. \ref{Fr-N6}. One
observes that the truncated equations work fine in the IR and produce the
predicted central charges. However, the iterations become unstable (the
pseudoenergies and also 
$F$ diverge to minus infinity) whenever $r$ is below some
critical ($N$-dependent) value. One finds moreover that $r_{crit}$ increases with $N$. 

\begin{figure}[H]
  \centering
\psfrag{rcrit}{$r_{crit}$}
  \includegraphics{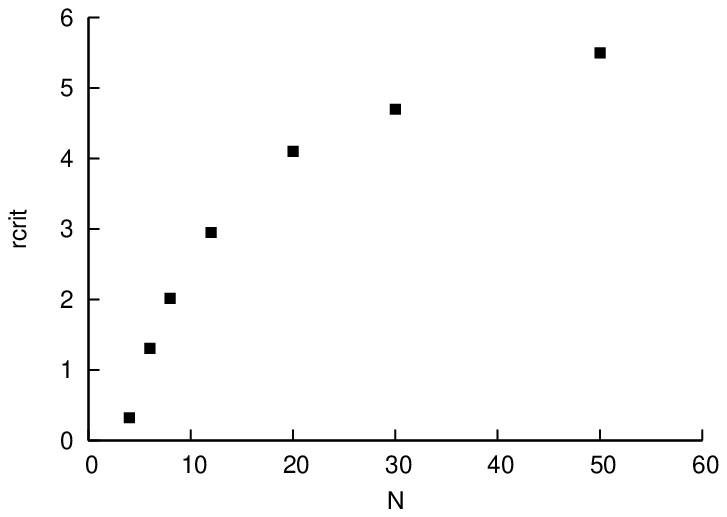}
\caption{Critical value of $r$ as a function of $N$}
\label{critr}
\end{figure}

We determined  $r_{crit}$ 
 for different values of
$N$, the results are shown in Table 1. and Fig. \ref{critr}. 
The numerical iterations become more
time demanding as $N$ is increased, therefore we had to content
ourselves with an accuracy of  $\Delta r=0.1$  for the larger values
of $N$.

\begin{table}[H]
\centering
    \begin{tabular}{|c|c|}
\hline
$N$ & $r_{crit}$ \\
\hline
4& 0.32175 \\
\hline
6& 1.305\\
\hline
8& 2.016 \\
\hline
12&  2.95\\
\hline
20& 4.1\\
\hline
30& 4.7\\
\hline
50& 5.5\\
\hline
    \end{tabular}
\caption{$r_{crit}$ for some values of $N$}
\end{table}

At this point  it is not clear what happens with $r_{crit}$ as $N$ is increased
further. There are two possibilities:
\begin{enumerate}
\item There exists a limiting value for $r_{crit}$ as $N\to
  \infty$. In this case the infinite set of TBA equations would be
  meaningful in the interval $r=r_{crit}\dots\infty$. However, one would see
  unusually high effective central charges, since already for $N=20$
  and $r=4$ one has $c_{eff}\approx 10$.
\item $r_{crit}\to\infty$ as $N\to\infty$.
\end{enumerate}

\begin{figure}[H]
  \centering
  \includegraphics{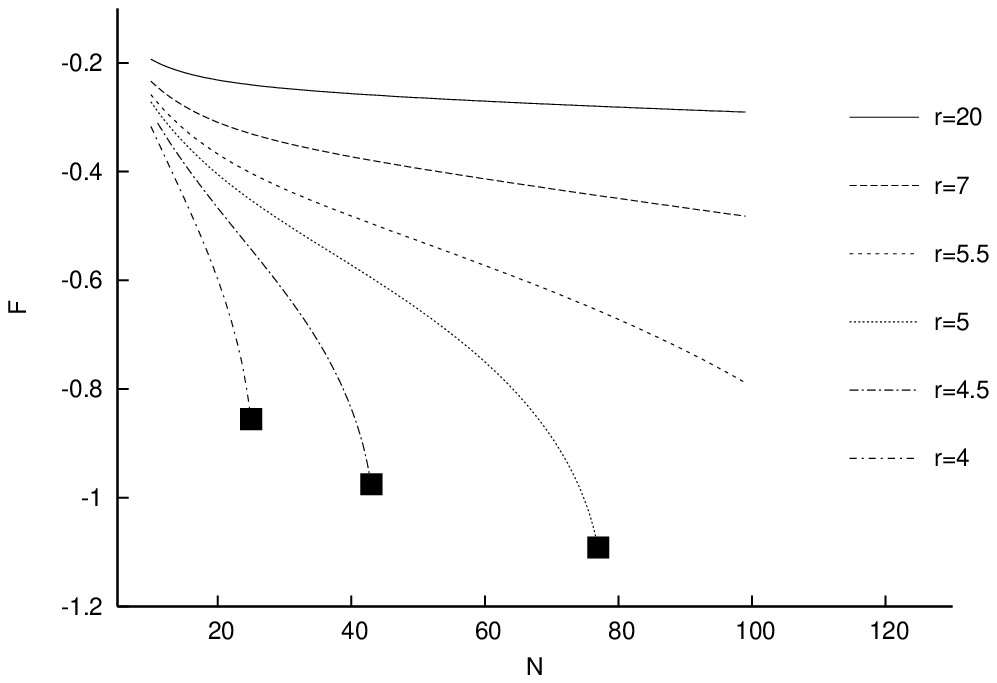}
\caption{$F$ as a function of $N$ for different fixed values of
  $r$. The solid square represent the critical values of $N$, where the
  iterations become unstable. }
\label{FN}
\end{figure}

To gain more insight, one can consider $F$ as a
function of $N$ for different fixed values of $r$; the results are
shown in Fig. \ref{FN}.  For $r\le 5$ we observed, that the iterations
become unstable whenever $N$ reaches a critical value $N_c<100$. These points are
marked with solid squares in the figure. For $r>5$ we did not find the
critical values of $N$, because we had to restrict the calculations to
$N<100$. However, from the figure it is clear that even for $r>5$ there is no
convergence. In fact one observes a linear gorwth in $N$ for every
value of $r$, suggesting that even for $r>5$
there always exists a (possibly quite large) critical $N$. This means
in turn, that in the $N\to\infty$ limit one has $r_{crit}\to\infty$.

We tried the improve the TBA by introducing alternative truncation
schemes for the infinite diagram, for example putting a ``fork'' at
the end. However, none of these tricks works to stabilize the
iterations for small $r$, because $\lambda_{\text max}$ will still be
greater than 2.

Based on the evidence presented above, we conclude that the infinite set
of equations \eqref{eq:massless_osp1} cannot be solved for any finite $r$.

\section*{Appendix B: Modifications of the  massless TBA}
\renewcommand\thesection{B}
\renewcommand{\theequation}{B\arabic{equation}}
\setcounter{equation}{0}
\setcounter{subsection}{0}

\subsection{CDD-factors}

It is a natural idea to add CDD factors to the conjectured S-matrix
and to consider the resulting TBA equations. 
CDD factors in
the LL and RR channels would alter the IR central
charge, and therefore we can exclude them. The only possibility is to introduce CDD factors for
$S_{RL}$ and $S_{LR}$. 

A simple factor of
\begin{equation*}
  f_{\alpha}=\frac{\sinh\theta-i\sin \alpha \pi}{\sinh\theta+i\sin \alpha \pi}
\end{equation*}
introduces a coupling between the massless nodes in the TBA equations
\begin{equation}
\begin{split}
  \eps_0&=r e^\theta/2-2s \star
  \ln\big(1+e^{-\eps_1}\big)-\varphi_{\alpha}\star\ln\big(1+e^{-\eps_{\bar{0}}}\big)  \\
 \eps_{\bar{0}}&=r e^{-\theta}/2-2s \star \ln\big(1+e^{-\eps_1}\big)
-\varphi_{\alpha}\star\ln\big(1+e^{-\eps_{0}}\big)\\
 \eps_1&=-s \star \ln\big(1+e^{-\eps_0}\big)-
s \star \ln\big(1+e^{-\eps_{\bar{0}}}\big) 
-s \star\ln\big(1+e^{-\eps_2}\big)\\
 \eps_i&-s \star \ln\big(1+e^{-\eps_{i-1}}\big)-
s \star \ln\big(1+e^{-\eps_{i+1}}\big) \quad \text{for} \quad
i=2,3,\dots
\end{split}
\label{simplest_CDD}
\end{equation}
where \begin{equation*}
  \varphi_{\alpha}=\frac{1}{2\pi}\frac{2\cosh\theta \sin \alpha\pi}{\sinh\theta^2+\sin^2 \alpha \pi}
\end{equation*}
The function $\varphi_{\alpha}$ has the same sign as
$s=\frac{1}{2\pi \cosh\theta}$ if $\alpha>0$, therefore it makes the
situation even worse. The only possibility to ''cure'' the TBA is to
choose $\alpha<0$. In this case there is a pole in the physical sheet
at $\theta=i\alpha$, which should be explained. However, the TBA makes
sense perfectly,
irrespective of the value of $\alpha$. In fact, $\alpha$ can be chosen
arbitrarily, because it only determines the crossover scale
$M$. Choosing $\alpha=1/2$ one obtains
\begin{equation*}
  f=\tanh^2(\theta/2+i\pi/4)\qquad\text{with}\qquad\varphi=-\frac{1}{\pi \cosh\theta}
\end{equation*}

The folded version of the TBA is then represented by the graph below,
where the vertical line with the arrows represents ,,negative'' coupling. 

\bigskip

\centerline{\includegraphics[scale=0.5]{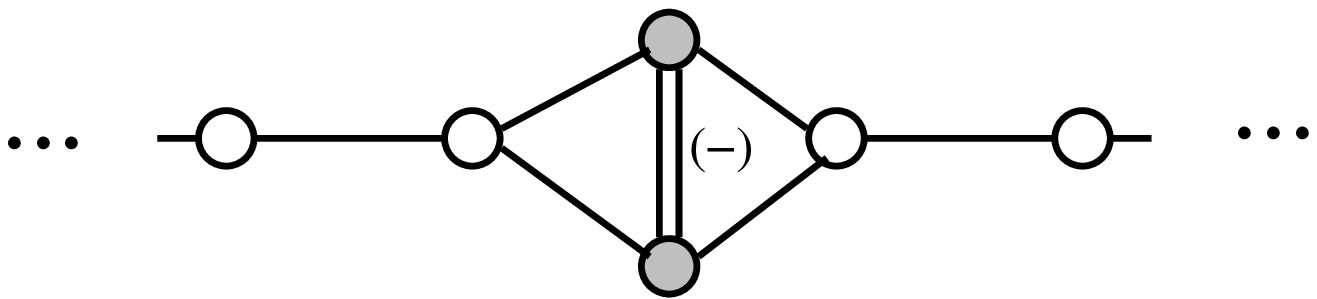}}

\bigskip

The UV central charge cannot be obtained in the usual way, because
there are still no finite plateus in $L=\log(1+e^{-\eps})$ and in
the central region one has $\eps_i=-\infty$ for every node. Instead
one can consider the truncated diagrams with $N$ nodes and let
$N\to\infty$. This way one obtains
\begin{equation*}
  c_{\hbox{eff}}=1\quad \text{(UV)}\qquad\qquad c_{\hbox{eff}}=3\quad \text{(IR)}
\end{equation*}
The above results were obtained by numerically solving the  Y-system
equations 
\begin{equation}
  \begin{split}
Y_0(\theta-i\pi/2)Y_0(\theta+i\pi/2)&=(1+Y_1(\theta))^2\frac{1}{(1+Y_{\bar 0}(\theta))^2}\\
Y_{\bar 0}(\theta-i\pi/2)Y_{\bar 0}(\theta+i\pi/2)&=(1+Y_1(\theta))^2\frac{1}{(1+Y_{0}(\theta))^2}\\
Y_1(\theta-i\pi/2)Y_1(\theta+i\pi/2)&=(1+Y_0(\theta))(1+Y_{\bar
  0}(\theta))(1+Y_2(\theta))\\
Y_i(\theta-i\pi/2)Y_i(\theta+i\pi/2)&=(1+Y_{i-1}(\theta))(1+Y_{i+1}(\theta))
  \end{split}
\end{equation}
with constant numbers $y_i$ and evaluating the central charge with the usual rules. We do
not know of an exact analytical proof. However, we can give an 
approximate solution, as follows. 

Let us consider a truncated diagram with $N$ nodes denoted
by $0,\bar{0},1,\dots (N-2)$. In the UV limit the variables
$y_i=e^{-\eps_i}$ take the values 
\begin{equation}
\label{x1}
  y_0=y_{\bar 0}\approx \sqrt{\frac{27(N-2)^2}{4\pi^2}}\qquad\qquad
y_i\approx\frac{9(N-2)^2}{2\pi^2}\left(1-\cos\left(\frac{2\pi (N-2-i)}{3(N-2)}\right)\right)
\end{equation}
in the central region. On the other hand, for $\theta\to\infty$ the
constant values are approximately
\begin{equation}
\label{x2}
  y_0=0\qquad
y_{\bar 0}\approx\frac{4(N-2)^2}{\pi^2}\qquad
y_i\approx\frac{2(N-2)^2}{\pi^2}\left(1-\cos\left(\frac{\pi (N-2-i)}{N-2}\right)\right)
\end{equation}
Actually, in this latter case the exact solution is known,
because this corresponds to the Y-system equations of the $A_n$
diagram with $n=2N-3$. However, the exact solution is not needed. The
effective central charge only depends on the way the variables $y_i$
approach infinity for $N\to\infty$ in \eqref{x1} and
\eqref{x2}. Substituting the above values into eq. (5.3) of \cite{dynkinTBA}
yields $c=1$ in the $N\to\infty$ limit.

\begin{figure}
  \centering
\psfrag{log10r}{$\log_{10} r$}
\psfrag{ceff}{$c_{eff}$}
\includegraphics{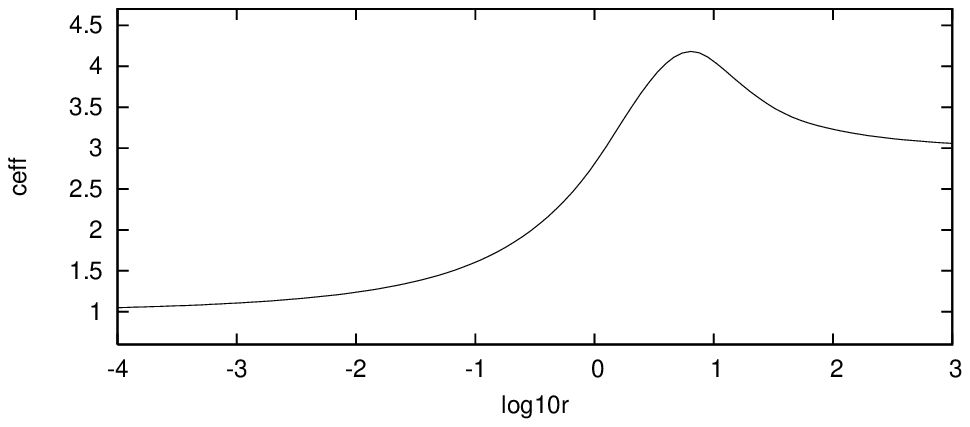}
  \caption{The evolution of the effective central charge in the
    massless TBA with a CDD factor.}
\label{fig:Ninfty_CDD}
\end{figure}

We performed numerical simulations of the truncated equations
\eqref{simplest_CDD} with different values of $N$ between $N=10$ and
$N=40$ and extrapolated the results to $N\to\infty$ applying a
polynomial fit in $1/N$. The results are shown in
Fig. \ref{fig:Ninfty_CDD}. 

We note that a simple factor of $\tanh(\theta/2+i\pi/4)$ is not sufficient to
stabilize the TBA, because in this case one still has  $\lambda> 2$
even for the truncated diagrams.

\subsection{$S_{LR}$ being a pure CDD factor}

In this case  (which was in fact discussed earlier in the literature \cite{Giuseppe})
the theory posseses two independent $OSP(2|2)$ 
symmetries corresponding to $LL$ and $RR$ scattering.The TBA we
are dealing with is now  represented by the diagram 

\bigskip

\centerline{ \includegraphics[scale=0.5]{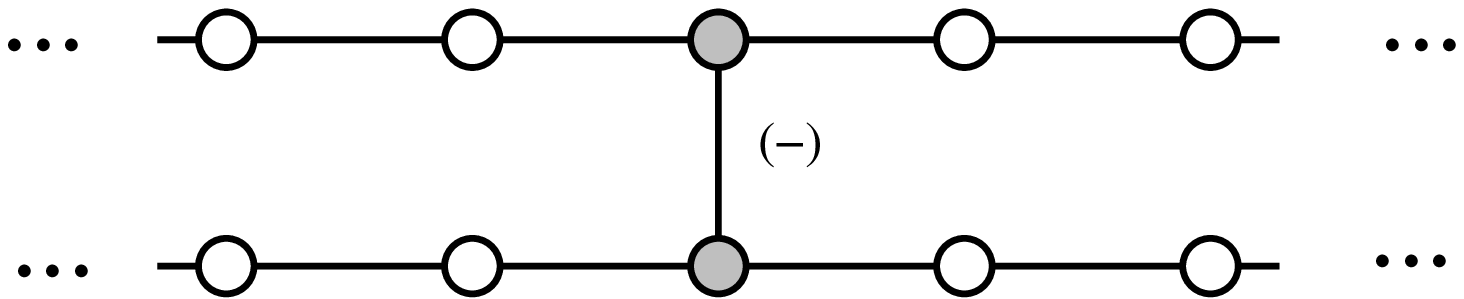}}

\bigskip

The vertical line connecting the two inifinite chains corresponds to the CDD
factor. A simple factor of
\begin{equation*}
  S_{LR}=\tanh(\theta/2-i\pi/4)
\end{equation*}
(which proved to be the correct choice in \cite{massless_Ising_flow} and \cite{massless_SU2})
does not work in this case, because once again the TBA will be
oversaturated. The simplest possibility is to choose the inverse of the
latter:
\begin{equation*}
  S_{LR}=\tanh(\theta/2+i\pi/4)
\end{equation*}
In this case the dashed link corresponds to 
\begin{equation*}
  \varphi=-\frac{1}{2\pi \cosh\theta}
\end{equation*}
The TBA  now  gives $c_{eff}=0$ in the UV (the truncated
diagrams have finite positive $c_{eff}$, which tends to zero if
$N\to\infty$), which is a probably meaningless result.

\end{document}